\documentstyle[prd,aps,epsf]{revtex}
\tightenlines

\newcommand{\aMSbar}{\alpha_{\overline{MS}}}
\let\orighspace\hspace
\renewcommand{\hspace}{\vrule width0pt\relax\orighspace}
\newcommand{\delsq}{\Delta^{(2)}}
\newcommand{\fBs}{f_{B_s}}

\def\lsim{\raise0.3ex\hbox{$<$\kern-0.75em\raise-1.1ex\hbox{$\sim$}}}
\def\gsim{\raise0.3ex\hbox{$>$\kern-0.75em\raise-1.1ex\hbox{$\sim$}}}

\begin{document}

\title{ 
{\normalsize \hfill {\sf UTCCP-P-99}} \\
\vspace*{-2pt}
{\normalsize \hfill {\sf March, 2001}} \\
  $B$ meson decay constants from two-flavor lattice QCD with
  non-relativistic heavy quarks
  }

\author{
  A.~Ali Khan$^1$\thanks{address till 31 August, 2000}, 
  S.~Aoki$^2$,
  R.~Burkhalter$^{1,2}$,
  S.~Ejiri$^1$, 
  M.~Fukugita$^3$, 
  S.~Hashimoto$^4$, 
  N.~Ishizuka$^{1,2}$,
  Y.~Iwasaki$^{1,2}$, 
  K.~Kanaya$^{1,2}$, 
  T.~Kaneko$^4$, 
  Y.~Kuramashi$^4$,
  T.~Manke$^1$\thanks{present address :
        Department of Physics, Columbia University, 
        New York, NY 10027. U. S. A.},
  K.~Nagai$^1$, 
  M.~Okawa$^4$, 
  H.P.~Shanahan$^1$\thanks{present address :
        Department of Biochemistry and Molecular
        Biology, University College London, London, England, UK}, 
  A.~Ukawa$^{1,2}$ and 
  T.~Yoshi\'e$^{1,2}$ \\
  (CP-PACS Collaboration) 
  }
\address{
  $^1$
  Center for Computational Physics,
  University of Tsukuba, Tsukuba, Ibaraki 305-8577, Japan;\\
  $^2$
  Institute of Physics, University of
  Tsukuba, Tsukuba, Ibaraki 305-8571, Japan;\\
  $^3$
  Institute for Cosmic Ray Research,
  University of Tokyo, Kashiwa, Chiba 277-8582, Japan; \\
  $^4$
  High Energy Accelerator Research Organization (KEK), 
  Tsukuba, Ibaraki 305-0801, Japan.
  }

\date{March 19, 2001}
\maketitle

\begin{abstract}
  We present a study of leptonic $B$ meson decay constants
  in lattice QCD with two flavors ($N_f=2$) of light
  dynamical quarks using NRQCD for the heavy quark.
  Gauge configurations are generated with a renormalization-group
  improved gauge action and a meanfield-improved clover
  light quark action.
  Measurements are carried out at two values of
  $\beta=6/g^2$, each for four sea quark masses,
  corresponding to the inverse lattice spacing
  $a^{-1}\approx 1.3$ and $1.8$ GeV in the chiral limit of
  sea quark. The continuum values of the decay constants are
  derived by evaluating the discretization errors at each finite
  lattice spacing. 
  We find $f_B^{N_f=2}=204(8)(29)(+44) $ MeV, 
  $f_{B_s}^{N_f=2} = 242(9)(34)(+38)$ MeV, and 
  $f_{B_s}^{N_f=2}/f_B^{N_f=2} = 1.179(18)(23)$, 
  where the errors listed are statistical, systematic and
  uncertainty due to choice of the physical quantity used to fix the
  scale.
  Comparison is made to quenched results ($N_f=0$) obtained
  with the same action combination and matching lattice
  spacings.
  We find 
  $f_B^{N_f=2}/f_B^{N_f=0}=1.07(5)$, 
  $f_{B_s}^{N_f=2}/f_{B_s}^{N_f=0}=1.10(5)$ and 
  $(f_{B_s}/f_B)^{N_f=2}/(f_{B_s}/f_B)^{N_f=0}=1.03(2)$, 
  which indicates a 5--10\% increase in the values of the
  decay constants, but no appreciable change in the ratio
  $f_{B_s}/f_B$, due to sea quarks.
\end{abstract}

\section{Introduction}
\label{sec:intro}

An accurate determination of the $B$ meson decay constant has 
practical importance, as it is needed, together with the
$\bar{B}-B$ transition matrix elements, for the
extraction of the quark mixing matrix elements from the $\bar{B}-B$
mass difference.
To this end much effort has been paid towards
a quantitative evaluation of these matrix elements using lattice
QCD (for reviews, 
see \cite{hashimoto99,bernard00}).

A problem specific to heavy meson calculations with the original 
lattice formulation is that the
heavy quark mass is greater than unity in units of lattice spacing,
which makes the lattice artifacts intolerably large. 
There are two popular formalisms to handle heavy quarks,
using the fact that the heavy quark in a $B$ meson is 
non-relativistic, in a way that large discretization effects 
proportional to the heavy quark mass do not appear.
One is the
use of the relativistic action with the aid of the non-relativistic
reinterpretation (the Fermilab formalism)~\cite{elkhadra97}, and
the other is simulations with the
NRQCD action~\cite{cornell92}. 

The problem with these formalisms, however, is 
that a continuum extrapolation is not simple.
For instance, since the NRQCD action is not renormalizable, an infinite
number of terms are involved in the $aM\rightarrow 0$ limit,
and their matching to the continuum theory requires good control
of power divergence of the form $1/(aM)^n$ with $n$ a
positive integer.
In the Fermilab formalism the continuum extrapolation
is possible in principle, although it is difficult in
practice because of a complicated $a$ dependence
of the couplings in the effective Hamiltonian of the heavy 
quark. 
Therefore, estimation of systematic errors arising from the
continuum extrapolation is a non-trivial task in both 
formalisms. 

Another uncertainty in previous calculations of heavy mesons 
is the effect of dynamical quarks. 
Most lattice studies to date adopt the quenched approximation. 
There are two calculations in which dynamical quarks are 
incorporated (full QCD) \cite{sara99,MILC99}. The results
indicate that the inclusion of dynamical quarks 
increases the value of $f_B$ and $\fBs$.  
The study by 
Collins {\it et al.}~\cite{sara99}, using NRQCD for heavy
quark and the clover action for light valence quarks, however, 
has been made only at a single lattice spacing in a small
physical volume, and with the Kogut-Susskind staggered 
action for sea quarks with a mass around that of the strange quark.
The other study by the MILC Collaboration~\cite{MILC99}, 
while covering a range of lattice spacings and sea quark
masses, uses the Wilson action both for heavy and
light valence quarks, but employs the staggered action for
sea quarks.  The problem with these calculations are that
the actions for valence and sea quarks are different, having
different symmetry structures at finite lattice spacings. This
would be an additional source of systematic errors. 

We study the $B$ meson decay constants, $f_B$ and $\fBs$,
incorporating
two flavors of light dynamical quarks, which are identified as
$u$ and $d$ quarks. The 
strange quark is treated in the quenched approximation. 
In order to test the estimation of systematic errors in the final 
results in the continuum limit, we compare the results from 
the two heavy quark formalisms.  
The work with the Fermilab formalism has been published
separately \cite{hugh}.
In this paper, we present the calculation using NRQCD.

We carry out full QCD calculation with a consistent use of
a quark action for both sea and light valence quarks. 
We adopt a renormalization group improved gauge
action~\cite{iwasaki85} for the gluon sector and an $O(a)$-improved 
clover action~\cite{clover85} for sea and light valence quarks 
\cite{burkhalter98,lightquark00,lighthadron00}.
The use of the improved action enables us to reduce
discretization errors, and makes it possible to study
decay constants at moderate lattice spacing, which is
feasible with the present computer resource.

The full QCD calculation is performed at two lattice
spacings ($1/a\simeq$ 1.3 and 1.8 GeV). 
In order to study the dynamical quark effects, we perform 
parallel quenched simulations with the same action, 
tuning the lattice spacing to be the same as those in full 
QCD simulations. 

The outline of this paper is as follows. 
In Sec.~\ref{sec:formalism}, we introduce the lattice
formulations of NRQCD used in this calculation, and
define the operators that contribute to the axial
vector current.
Simulation details, such as the run parameters and choice of
operators, and fitting methods are described in
Sec.~\ref{sec:simulation}. 
In Sec.~\ref{sec:analysis}, we discuss determinations of 
the physical lattice spacing, the $B$ meson masses and 
decay constants. A detailed discussion on the
extrapolation and interpolation to the physical quark masses 
is given. 
We then attempt to obtain 
the continuum results by estimating the discretization errors
at each finite lattice spacing in Sec.~\ref{sec:continuum}. 
A comparison with results of previous studies is made in
Sec.~\ref{sec:comparison}, and the 
conclusions are given in Sec.~\ref{sec:conclusions}. 
Detailed numerical results are collected in the Appendix.

\section{Formalism}
\label{sec:formalism}
\subsection{Actions}
For gluons we adopt a renormalization-group (RG) improved
gauge action consisting of plaquettes and $1 \times 2$
rectangular Wilson loops \cite{iwasaki85}.
The action for both sea quarks and light valence quarks is 
taken to be the $O(a)$-improved clover
action~\cite{clover85} with a meanfield-improved clover 
coefficient $c_{SW} = P^{-3/4}$, where the plaquette $P$ is 
evaluated in one-loop perturbation theory as
$P=1-0.8412/\beta$. 
For our choice of parameters the measured values of
plaquette $\langle P\rangle$ is well approximated by the
one-loop evaluation \cite{lighthadron00}.

For the heavy quarks we use NRQCD corrected to
$O(1/M_0)$ with $M_0$ the heavy quark mass. 
Previous quenched NRQCD calculations have shown that
$O(1/M^2_0)$ corrections are small in the decay constants,
being of the order of $\sim 3-4\%$ at the lattice spacings
used in this study~\cite{ishikawa97,alikhan98,JLQCD00}.
There are various ways to discretize an
NRQCD action 
(see {\it e.g.}, \cite{cornell92,alikhan98,davies94}). 
Here, we choose the action that is symmetric under a time
reversal transformation:
\begin{equation}
  S_{NRQCD} = \sum
  \psi_t^{\dagger} \left[
    \psi_t - 
    \left( 1-\frac{a\delta H}{2} \right)_t
    \left( 1-\frac{aH_0}{2n} \right)_t^n 
    U^\dagger_4
    \left( 1-\frac{aH_0}{2n} \right)_{t-1}^n 
    \left( 1-\frac{a\delta H}{2} \right)_{t-1}
    \psi_{t-1}
  \right]
  \label{eq:evol},
\end{equation}
where $\psi_t$ is a two-component Pauli spinor at 
a time slice $t$.
The sum runs over all lattice sites, while indices to
represent spatial positions are suppressed.
The operators $H_0$ and $\delta H$ correspond to the
non-relativistic kinetic energy and the spin-chromomagnetic
interaction, as defined by 
\begin{eqnarray}
  H_0 & \equiv & - \frac{\delsq}{2M_0}, \\
  \delta H  & \equiv &
  - c_B \frac{g}{2M_0} \vec{\sigma}\cdot\vec{B},
\end{eqnarray}
respectively, 
where $\delsq$ is a Laplacian discretized in a standard way
and $\vec{B}$ represents a chromomagnetic field strength
defined with a clover-leaf shape as in \cite{cornell92}.
The stabilization parameter $n$ is introduced to avoid 
an instability of the Green function at large separation 
due to high-momentum modes. 
In the free theory, this parameter should satisfy 
$n > 3/(2aM_0)$~\cite{cornell92}.
We choose a larger value of $n$ to 
make the simulation stable~\cite{alikhan98,JLQCD00}. 
The actual numbers are given in Table~\ref{tab:para_dyn} for full
and in Table~\ref{tab:para_que} for quenched QCD
runs.
Gauge links appearing in the NRQCD action are
meanfield-improved, $U_{\mu} \rightarrow U_{\mu}/u_0$, with
$u_0$ determined from the mean link in Landau gauge.
The coefficient $c_B$ is set to its tree-level value, {\it i.e.} unity.
We let the heavy quark Green functions to evolve from $t$=0 to
$T/2$, and a time reversed evolution through the second half of
the lattice from the same source.

\subsection{Current operators}
\label{sec:formalism:currents}
The pseudoscalar decay constant is defined by 
\begin{equation}
  i f_B p_\mu = \langle 0|A_\mu|B(p)\rangle\ ,
\end{equation}
where the axial current 
$A_\mu = \bar{q}\gamma_5\gamma_{\mu}h$ is formed with 
the relativistic spinors $q$ for the
light quark and $h$ for the heavy quark.
We restrict our considerations to the
time component, $A_0$.

The heavy quark field $h$ is related to a non-relativistic
field $Q$ via a Foldy-Wouthuysen-Tani transformation at the tree
level. 
Ignoring the $O(1/M^2)$ terms, this reads
\begin{equation}
  \label{eq:FWT}
  h = 
  \left(
    1 - \frac{\vec{\gamma}\cdot\vec{\nabla}}{2M_0} 
  \right) Q,
\end{equation}
where
\begin{equation}
  Q = 
  \left( 
    \begin{array}{c}
      \psi \\ 0
    \end{array}
  \right).
\end{equation}
We write the two lattice operators contributing to the time
component of the heavy-light axial vector current as
\begin{eqnarray}
  J_L^{(0)} & = & \bar{q}\gamma_5\gamma_0 Q, \nonumber \\
  J_L^{(1)} & = & - \frac{1}{2M_0}
  \bar{q}\gamma_5\gamma_0\vec{\gamma}\cdot\vec{\nabla}Q.
\end{eqnarray}
In matching the continuum operator to the
lattice counterparts, an additional operator $J_L^{(2)}$
appears at $O(1/M)$
\begin{equation}
  J_L^{(2)} = \frac{1}{2M_0}
  \bar{q} \vec{\gamma}\cdot\stackrel{\leftarrow}{\nabla}
  \gamma_5\gamma_0 Q\ ,
\end{equation}
though its matrix element is equal to that of $J_L^{(1)}$ at
zero momentum due to translational invariance on the
lattice. 
Thus, at the one-loop level we use
\begin{equation}
  \label{eq:matching}
  A_0 = (1+\alpha_s\rho_A^{(0)}) J_L^{(0)}
      + (1+\alpha_s\rho_A^{(1)}) J_L^{(1)}
      + \alpha_s\rho_A^{(2)} J_L^{(2)},
\end{equation}
where $\rho_A^{(0)}$, $\rho_A^{(1)}$ and $\rho_A^{(2)}$ have been computed in 
Ref.~\cite{ishikawa99} 
\footnote{
 The definition of the lattice currents
  $J_4^{(i)}$ in \cite{ishikawa99} is slightly different
  from $J_L^{(i)}$ used in this paper.
  The tree level rotation is included in $J_4^{(0)}$ in
  \cite{ishikawa99}. 
  The one-loop coefficients $\rho_A^{(i)}$ are modified
  accordingly. 
  }
for the RG-improved gauge action.
The coefficient $\rho_A^{(2)}$ diverges in the limit of
$aM_0\rightarrow\infty$ as 0.34$\times 2aM_0$, which cancels
the factor $1/M_0$ in the definition of $J_L^{(2)}$ and
gives a finite contribution. 
This remaining contribution corresponds to the improvement
of discretization error of $O(\alpha_s a)$
\cite{morningstar-shigemitsu98} in the static limit.

For $\alpha_s$, we use the coupling defined with the
$\overline{MS}$ scheme at $\mu = 1/a$, which is
evaluated using the one-loop relation 
\begin{equation}
  \label{eq:alpha} 
  \frac{1}{g^2_{\overline{MS}}(\mu)} = 
  \frac{(c_0P-8c_1R) \beta}{6} -
  0.1006 + 
  \frac{22}{16\pi^2}\log{(\mu a)} + 
  N_f \left(
    0.03149 - \frac{4}{48\pi^2}\log{(\mu a)}
  \right).
\end{equation}
The meanfield improvement is applied using measured values of
 the $1\times 1$ and $1\times 2$ Wilson loops, $P$ and $R$.

\section{Simulation Details}
\label{sec:simulation}
\subsection{Gauge configurations}
The CP-PACS Collaboration has generated a set of full QCD gauge
configurations incorporating two flavors of light dynamical
quarks at four values of gauge coupling, $\beta=1.8$, 1.95, 2.1, 2.2, 
using the Hybrid Monte Carlo (HMC) algorithm 
\cite{burkhalter98,lightquark00,lighthadron00}.
At each $\beta$, four sea quark masses $m_{sea}$ are
chosen in the range $0.5m_s \lsim m_{sea} \lsim 3m_s$, 
where $m_s$ denotes the strange quark mass.
In the chiral limit of sea quarks, these couplings
correspond to lattice spacings $a$ with 
1 GeV $\lsim a^{-1} \lsim 2.2$ GeV.
The spatial lattice size is about 2.5 fm at $\beta=1.8$, 
1.95 and 2.1, while it is about 1.9 fm at $\beta=2.2$.
The parameters of the configurations
are summarized in the left half of Table~\ref{tab:param-all}.  

In the present study we use the configurations 
at $\beta=1.95$ ($a^{-1} \simeq 1.3$ GeV) and 
2.1 ($a^{-1} \simeq 1.8$ GeV);
the entire gauge configurations we generated are analyzed at 
$\beta=1.95$, whereas half
the configurations, corresponding to the first 2000
trajectories, are employed for $\beta=2.1$.
The configurations are separated by 10 HMC trajectories at
$\beta=1.95$ and by 5 trajectories at 2.1.
We discard configurations at $\beta=1.8$, because of 
expected large discretization errors, and at $\beta=2.2$
for a too small physical volume. 
Details of the parameters employed in our analysis are listed in
Table~\ref{tab:para_dyn}.

The CP-PACS study of full QCD is supplemented by quenched simulations. 
These simulations are based on a set of quenched
configurations with the same gauge action separated by 100
sweeps, each sweep consisting of 1 heatbath and 4 overrelaxation steps.
The gauge couplings in the quenched configurations are
tuned such that the string tension matches the one in full QCD  
at four sea quark masses in our simulation
and at the physical sea quark mass point corresponding $u$ and $d$ quarks, 
as shown in the right half of Table~\ref{tab:param-all} \cite{hugh}.
We analyze a subset of the quenched configurations at four
values of $\beta$ as listed in Table~\ref{tab:para_que}.
The first two values ($\beta=2.187$ and 2.281) correspond to
the full QCD runs with the heaviest and the lightest sea
quark masses at $\beta=1.95$.
The third ($\beta=2.334$) and the fourth, 
($\beta$=2.575), correspond to full QCD at
$\beta=1.95$ and 2.1 at the physical sea quark mass, respectively.
Details of the parameters of these quenched runs are given
in Table~\ref{tab:para_que}.

\subsection{Correlator measurements}

In order to improve the overlap of interpolating field with
the heavy-light meson ground state, we use smeared operators
$O_S$
\begin{equation}
  O_S(\vec{x},t) = 
  \sum_{\vec{y}}
  \bar{q}(\vec{x},t)\gamma_5 Q(\vec{y},t)
  \phi(|\vec{x}-\vec{y}|),
\end{equation}
on the gauge configurations fixed to the Coulomb gauge.
For the smearing function $\phi(|\vec{x}-\vec{y}|)$ we use
an exponential form,
\begin{eqnarray}
  \phi(|\vec{x}-\vec{y}|) & = & A \exp(-B|\vec{x}-\vec{y}|)
  \;\; {\rm for}\; |\vec{x}-\vec{y}| \neq 0
  \nonumber \\
  \phi(0) & = & 1.
  \label{eq:smeared}
\end{eqnarray}
We calculate correlators using smeared operators at the
source and using both  local and smeared
operators at the sink,
\begin{eqnarray}
  C_{SS}(t) & = & \sum_{\vec{x}} 
  \langle O_S(\vec{x},t) O^\dagger_S(\vec{0},0) \rangle,
  \nonumber \\
  C_{SL}(t,|\vec{p}|) & = & \sum_{\vec{x}}
  e^{i\vec{p}\cdot\vec{x}}
  \langle O_L(\vec{x},t) O^\dagger_S(\vec{0},0) \rangle,
\end{eqnarray}
where $O_L$ denotes a local pseudoscalar density.
The spatial momentum $\vec{p}$ is introduced for $C_{SL}$ to
study the energy-momentum dispersion relation of 
mesons. We make measurements for momenta $p^2$ = $|\vec{p}|^2$ =
0, 1, ..., 5 in units of $(2\pi/La)^2$ with $L$ the spatial
extent of the lattice.
We average over all possible spatial directions.

The correlators for the axial currents are given by
\begin{equation}
  C^{(i)}_{SL}(t) = \sum_{\vec{x}}
  \langle J_L^{(i)}(\vec{x},t) O_S^\dagger(\vec{0},0) \rangle,
  \qquad i=0,1,2. 
\end{equation}
Note that $C^{(0)}_{SL}(t)\equiv C_{SL}(t,\vec{0})$ since
the lower two components of $Q$ vanish. 

\subsection{Correlator fits}
To calculate the decay constants we need the amplitude
of the local currents.
Since smeared-smeared and smeared-local correlators both
have the exponential falloff with the same exponent controlled by
$E_{sim}$, we make a simultaneous fit to a single exponential as
\begin{eqnarray}
  \label{eq:C_SS}
  C_{SS}(t)       & = & Z_S^2 \exp{(-E_{sim}t)}, \\
  \label{eq:C_SL}
  C^{(i)}_{SL}(t) & = &
  Z^{(i)}_L Z_S \exp{(-E_{sim}t)},\qquad i=0,1,2. 
  \label{eq:ampls}
\end{eqnarray}

We apply  a bootstrap procedure with 500 samples, 
taking correlations between different correlators and time
slices into account. 
For the dynamical configurations, we bin over 2
configurations at $\beta$=1.95 and 5 configurations at
$\beta$=2.1. 
The quenched configurations are
regarded as independent and are not binned. 
Results from forward and time reversed evolution on the same
configurations are always averaged.
In the full QCD analysis at $\beta$=2.1 we introduce a
cutoff on the ratio of the largest to the smallest eigenvalues of
the covariance matrix to avoid a low value of the
goodness-of-fit $Q$. 
We have checked for the heaviest and the lightest sea quark mass
that the effect of this procedure on the decay constants is
small, amounting to at most $\sim$ 25\% of the
statistical error. 
The problem of a low $Q$ at $\beta$=2.1 might arise from the smaller
number of independent
configurations.

The fitting interval $[t_{min},t_{max}]$ is
chosen such that both correlators already reach a plateau at
$t_{min}$. 
This is judged by $Q$, as well as by eye from the
effective mass and effective amplitude plots. 
Correlators, in particular $C_{SS}$, are often dominated by
noise for large $t$, which is indicated by a decrease of  
$Q$ for a large $t_{max}$. 
We cut the fit at $t_{max}$ before noise dominates. 

\begin{figure}[tbp]  
  \begin{center}
    \centerline{
      \epsfysize=0.4\vsize
      \epsfbox{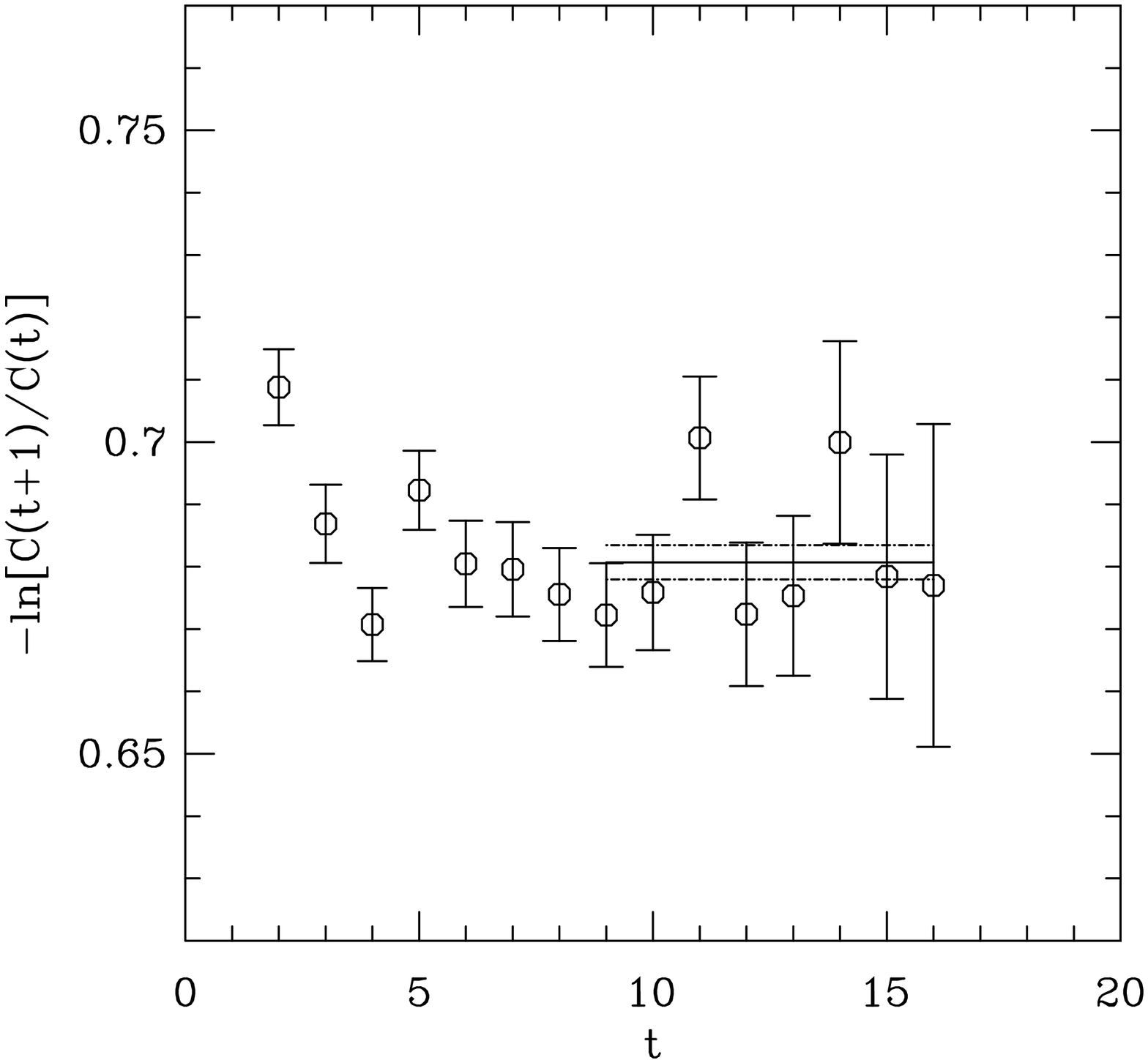} 
      \epsfysize=0.4\vsize
      \epsfbox{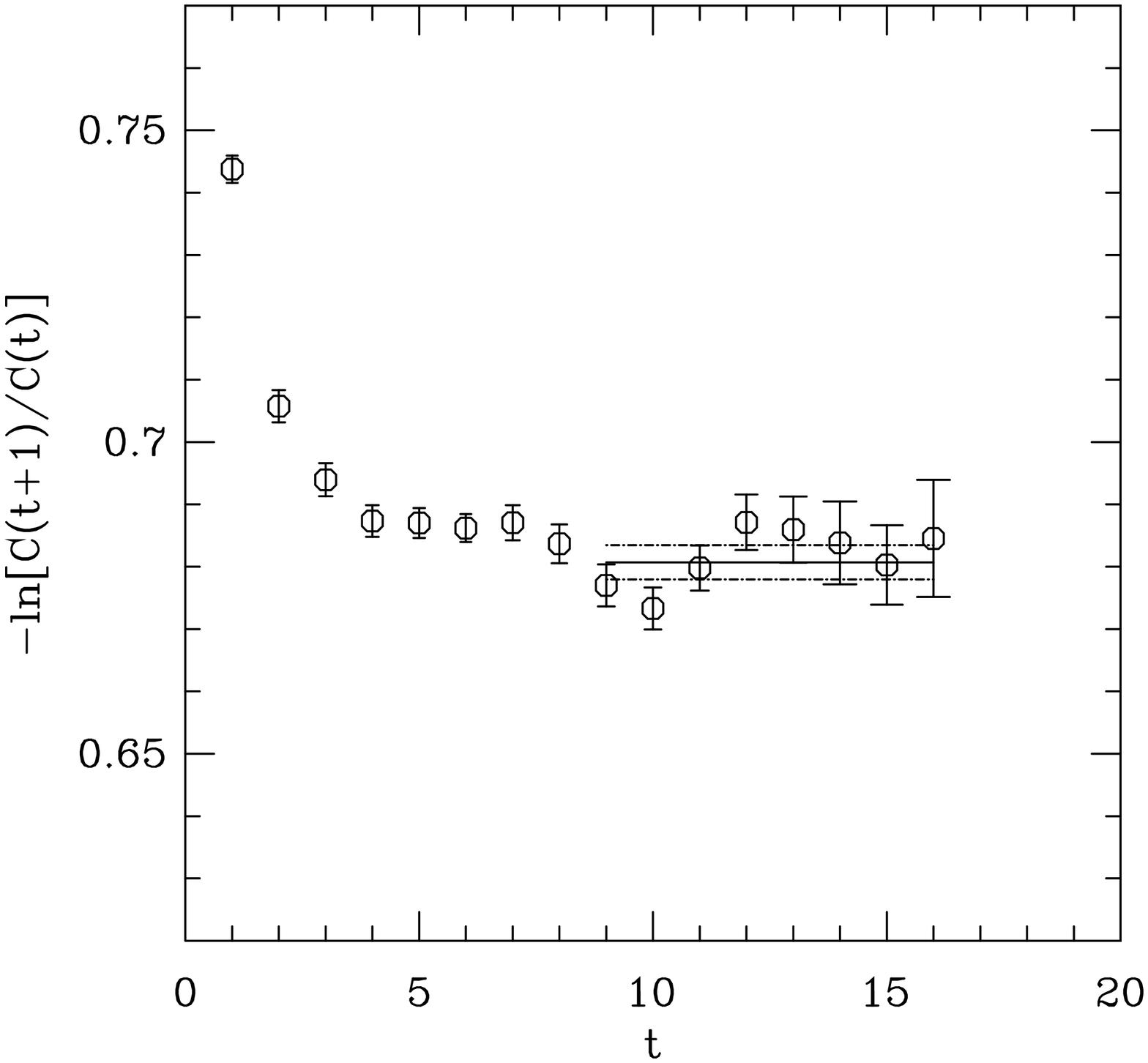}}
    \vspace{-1.3cm}
    \centerline{
      \epsfysize=0.4\vsize \epsfbox{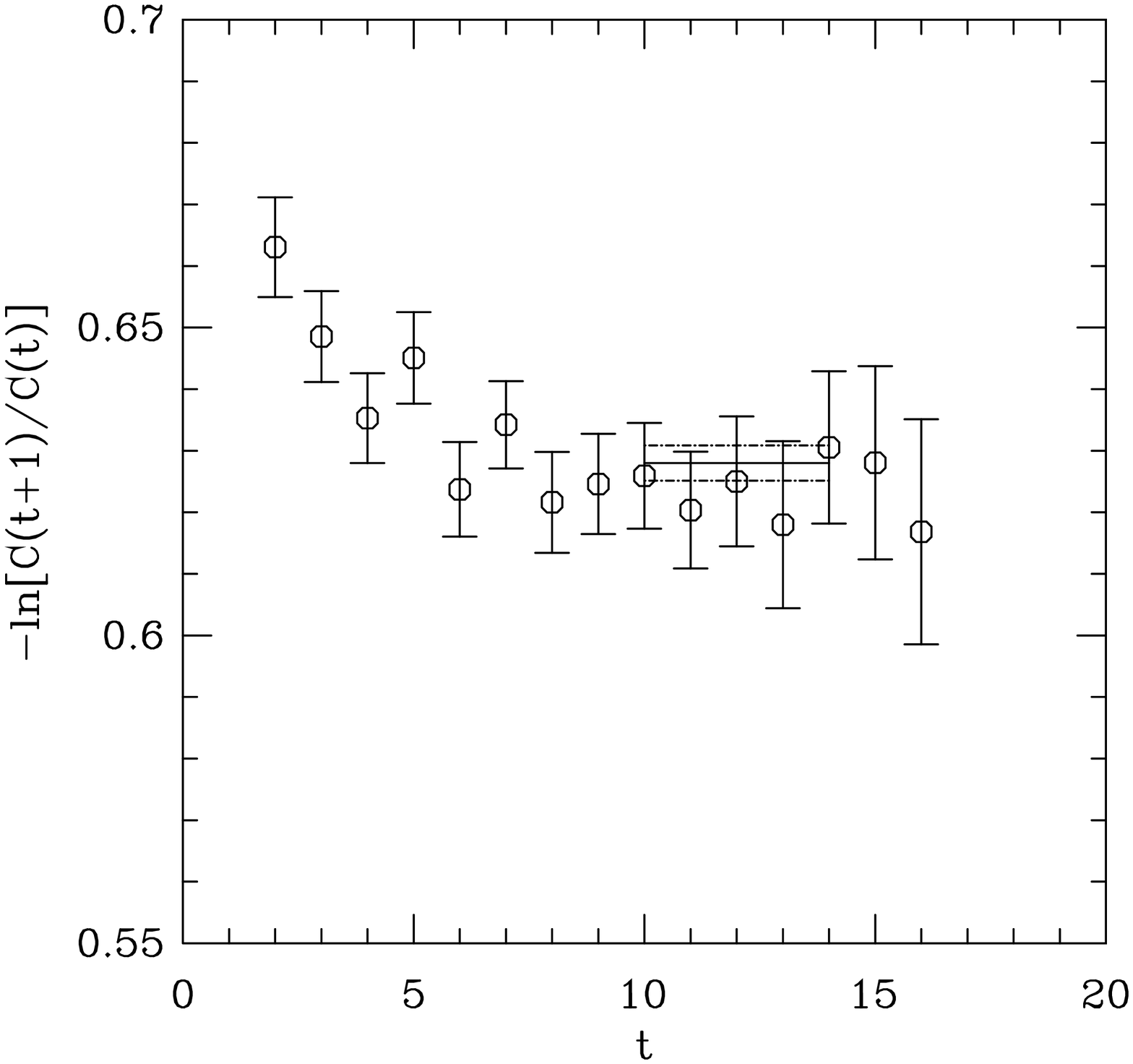}
      \epsfysize=0.4\vsize \epsfbox{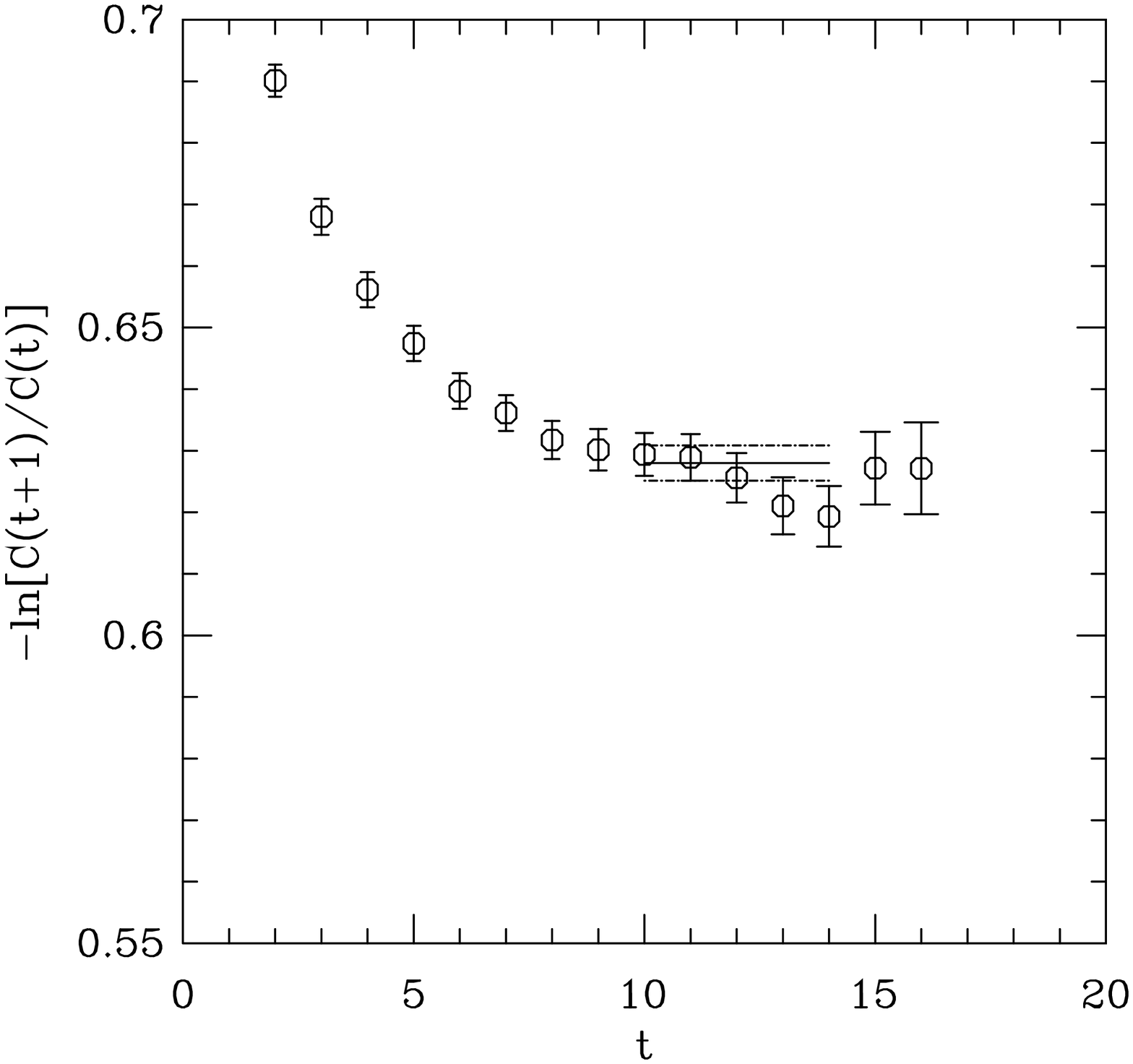}}
  \end{center}
  \vspace{-0.5cm}
  \caption{
    Upper panels show effective mass . of $C_{SS}$
    (left) and $C^{(0)}_{SL}$ (right) in full QCD at 
    $\beta$ = 1.95, $K_{sea}$ = 0.1410, $K_{val}$ = 0.1390,
    and $aM_0$ = 4.0.
    Lower panels are for quenched QCD at $\beta$ = 2.334
    for $K$ = 0.1349 and $aM_0$ = 4.0. 
    Solid and dashed lines show fit results and one standard 
    deviation error bands.
    } 
  \label{fig:effmass1.95}
\end{figure}

\begin{figure}[tbp]  
  \begin{center}
    \centerline{
      \epsfysize=0.4\vsize
      \epsfbox{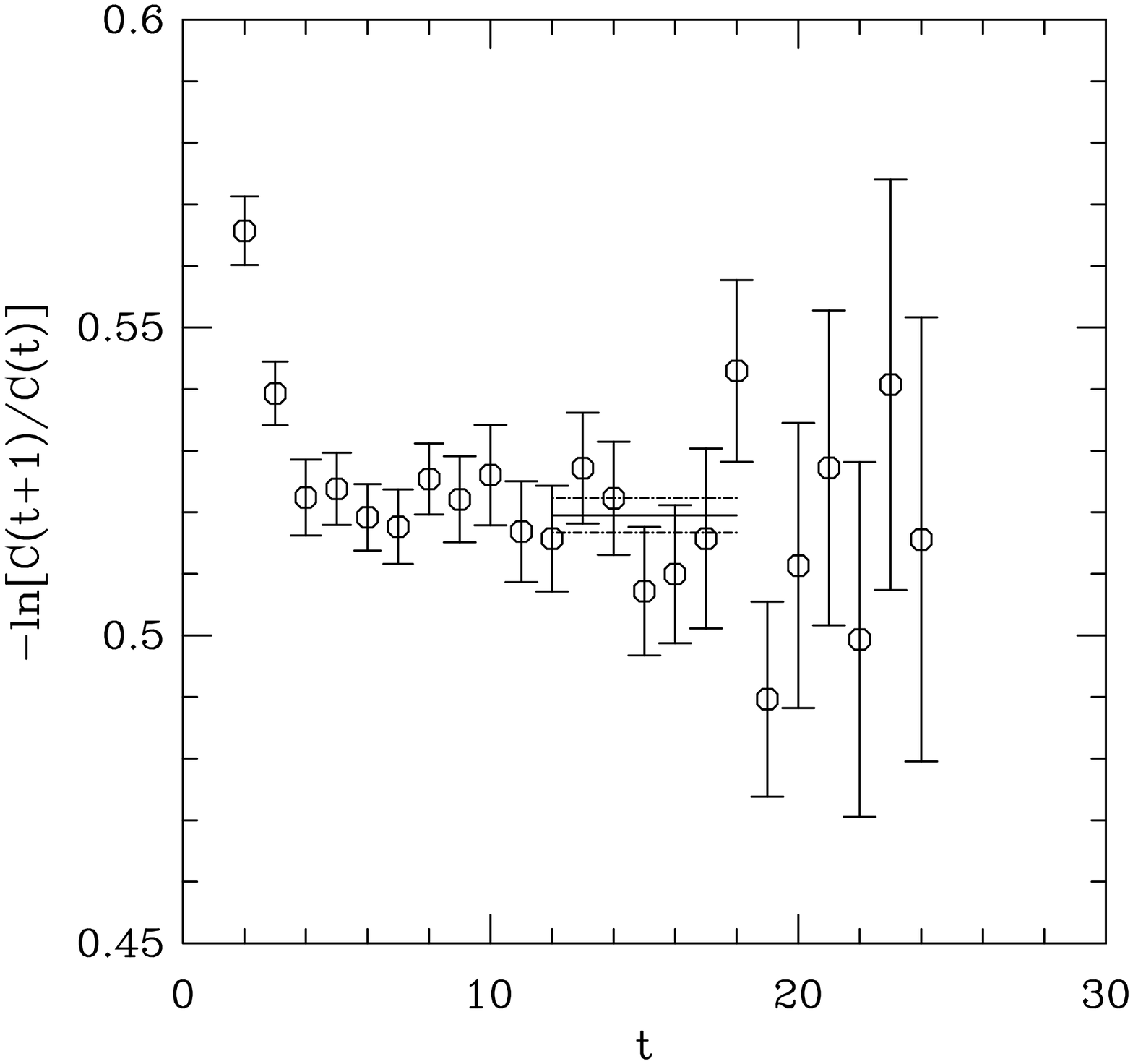} 
      \epsfysize=0.4\vsize
      \epsfbox{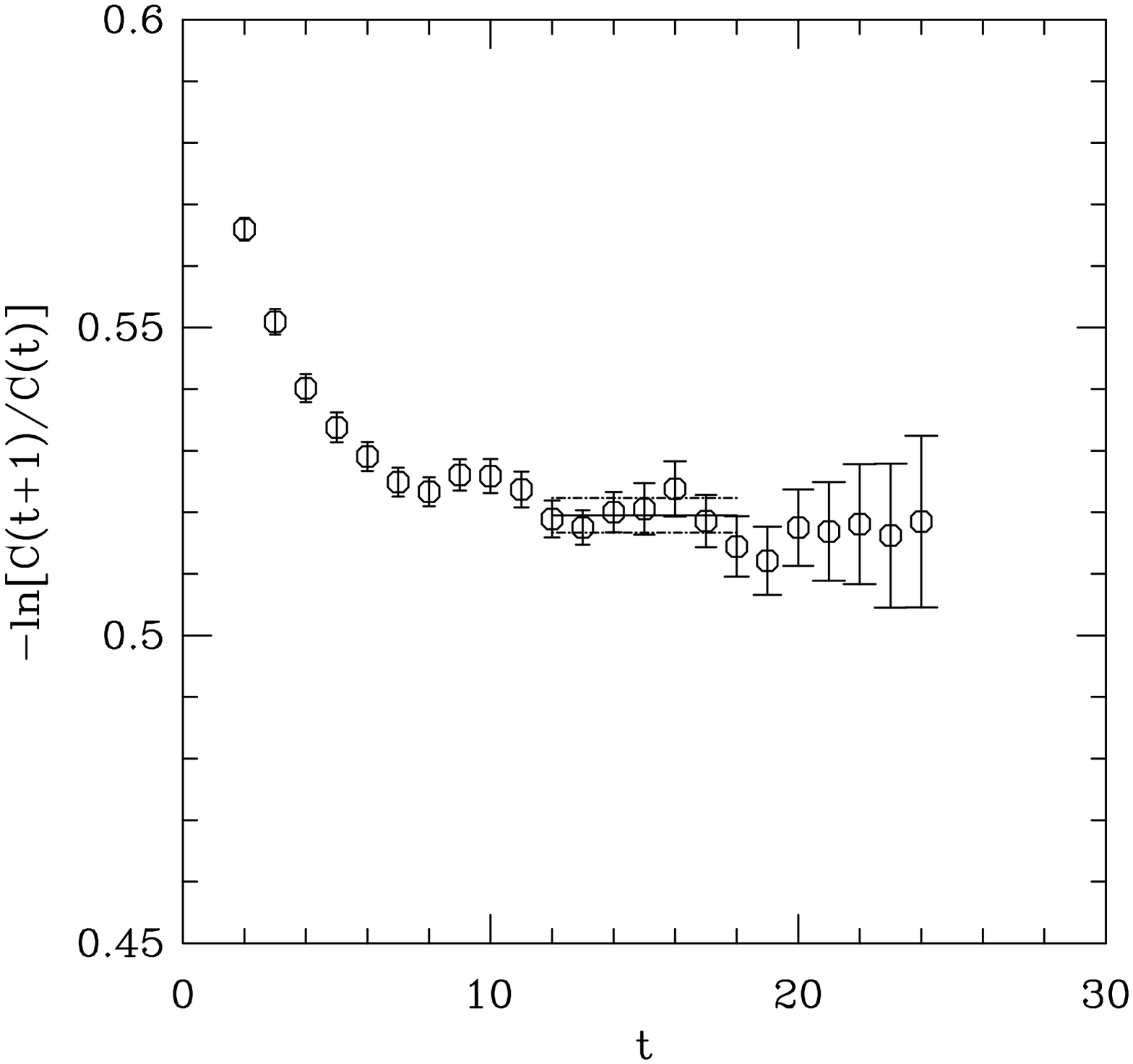}}
    \vspace{-1.3cm}
    \centerline{
      \epsfysize=0.4\vsize \epsfbox{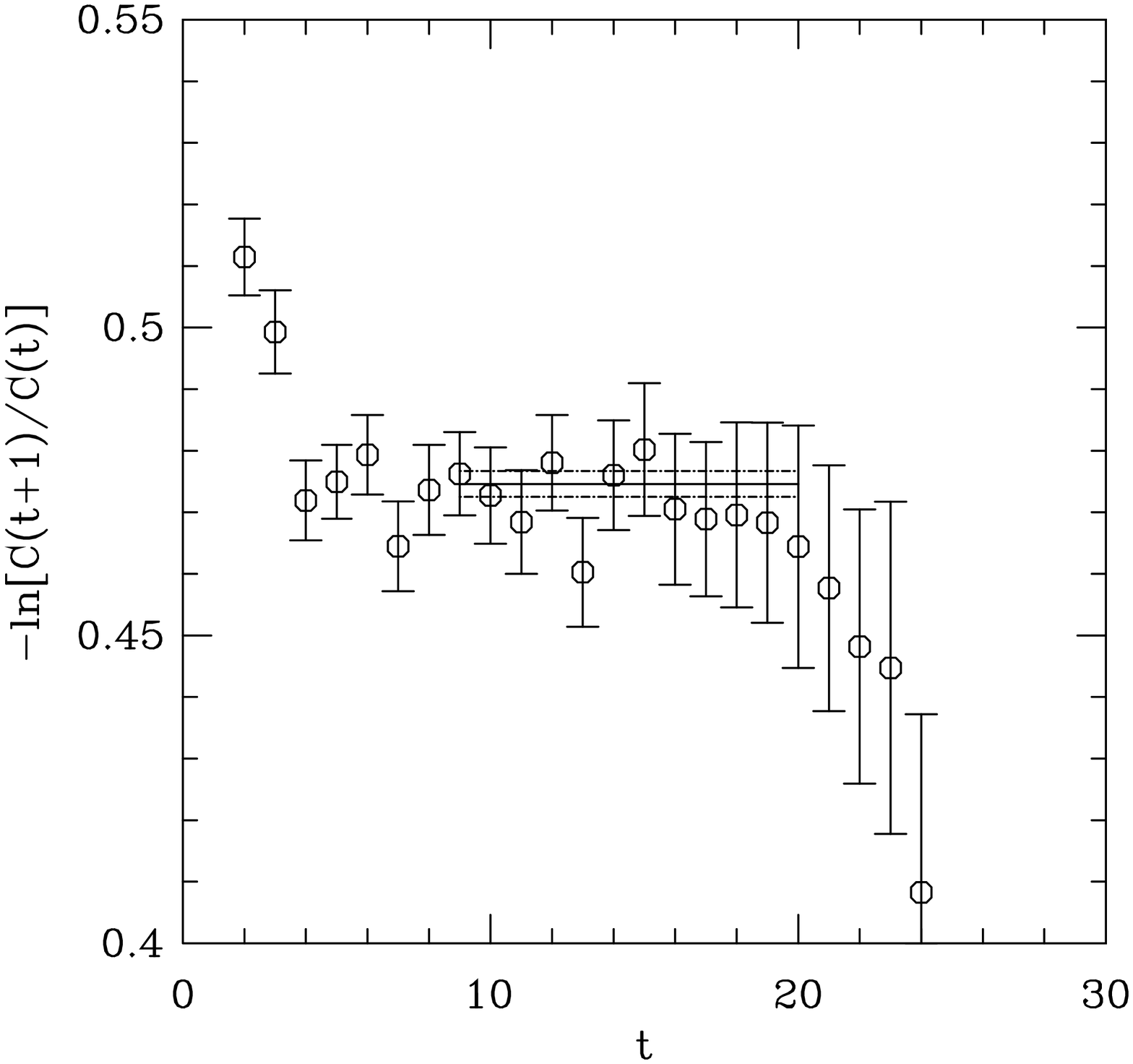}
      \epsfysize=0.4\vsize \epsfbox{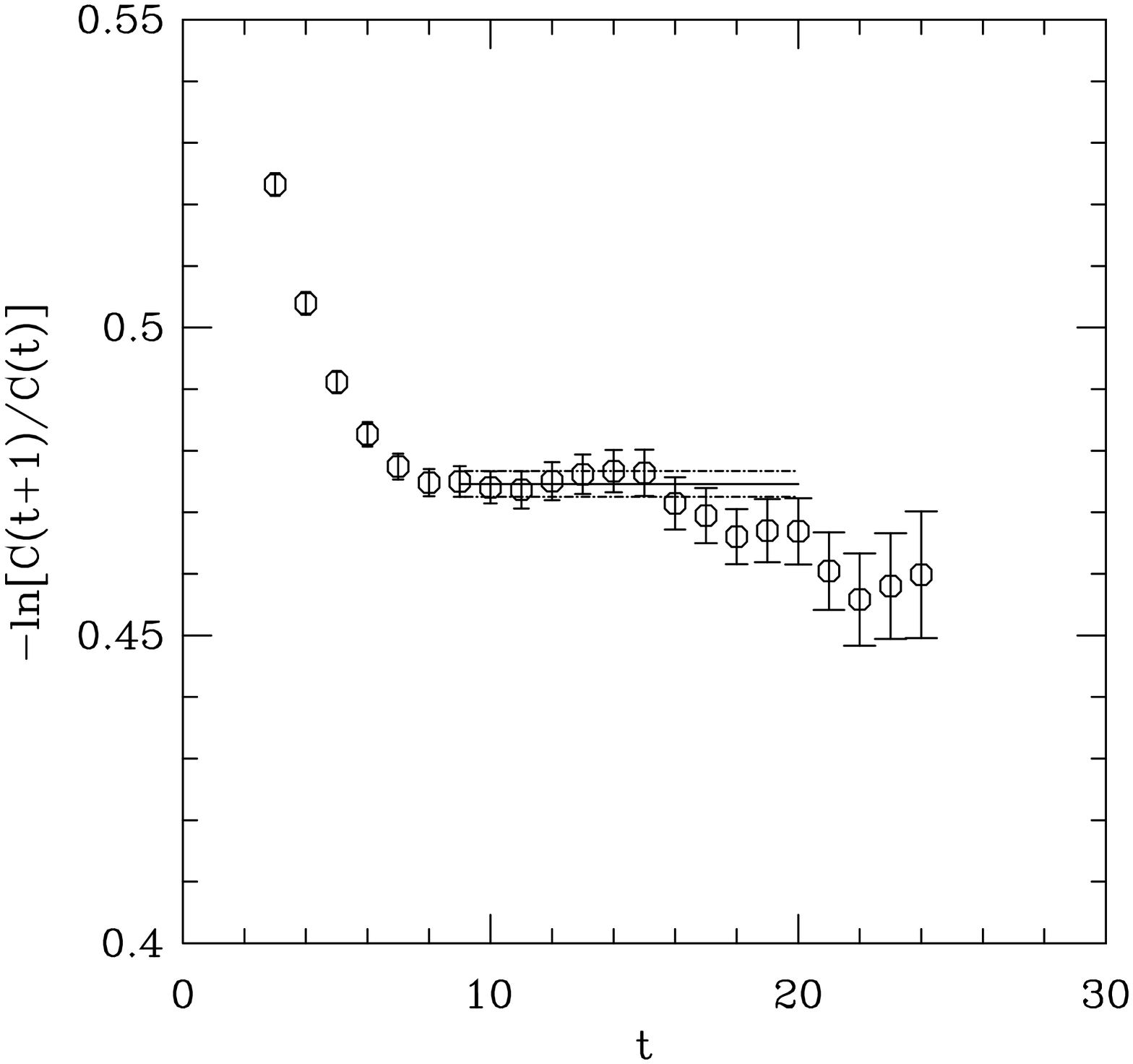}}
  \end{center}
  \vspace{-0.5cm}
  \caption{
    Upper panels show effective mass . of $C_{SS}$ (left)
    and $C^{(0)}_{SL}$ (right) in full QCD at $\beta$ = 2.1,
    $K_{sea}$ = 0.1382, $K_{val}$ = 0.1374, and $aM_0$ = 2.9.
    Lower panles are for quenched QCD at $\beta$ = 2.575,
    $K_{val}$ = 0.1374, and $aM_0$ = 2.9.
    } 
  \label{fig:effmass2.1}
\end{figure}

Examples of the effective mass plots for $C^{(0)}_{SL}$ and $C_{SS}$ 
are shown in Figs.~\ref{fig:effmass1.95} and
\ref{fig:effmass2.1}, where 
the light valence quark mass is taken approximately 
to be the strange quark mass 
and the heavy quark mass to be the $b$ quark.
In Fig.~\ref{fig:effmass1.95} the upper two panels
illustrate full QCD data at $\beta$=1.95 with the lightest
sea quark mass, and the lower two show quenched
results at $\beta$=2.334 corresponding to the physical value
of the sea quark mass in full QCD at $\beta$=1.95.
Figure~\ref{fig:effmass2.1} shows similar plots at
$\beta$=2.1.

\begin{figure}[tbp]  
  \begin{center}
    \epsfysize=0.4\vsize
    \epsfbox{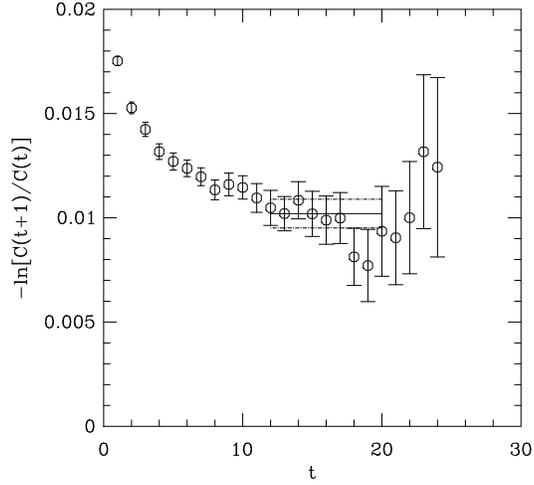}
  \end{center}
  \caption{
    Effective mass plot of the ratio of correlators at 
    $p = 2\pi/L$ and at $p$ = 0 for $\beta$ = 2.1,
    $K_{sea}$ = 0.1382.
    Valence quark masses correspond to $K_{val}$ = 0.1374 and
    $aM_0$ = 2.9.
    } 
  \label{fig:rffmass2.1}
\end{figure}

Energy for a given finite spatial momentum $E_{sim}(p^2)$ is 
extracted from the difference 
$\Delta E(p^2) \equiv E_{sim}(p^2) - E_{sim}(0)$
using a single exponential fit to the ratio 
\begin{equation}
  \frac{C_{SL}(t,|\vec{p}|)}{C_{SL}(t,|\vec{0}|)}
  = A(|\vec{p}|) \exp{(- \Delta E(p^2) t)}.
\end{equation}
An example of the effective mass is shown in
Fig.~\ref{fig:rffmass2.1} for full QCD simulations at
$\beta$=2.1.

\section{Analysis}
\label{sec:analysis}

\subsection{Fixing the physical scale}

We carry out a partially quenched analysis of
the dynamical configurations. 
Namely, we evaluate the decay constants for the
configurations at each sea quark mass, at which the
lattice spacing and physical quark masses are determined by
varying the valence light quark mass as we do in the
quenched analysis.
We then extrapolate the results to
the physical value of the sea quark mass.
In this subsection we discuss the determination of the lattice
scale.

In simulations of full QCD with the realistic spectrum of
dynamical sea quarks, one should in principle obtain a unique value for
the lattice spacing $a$ from any physical quantity it is determined.
In our simulation, however, we obtain different values of $a$
depending on the quantity used to fix the scale since dynamical quark 
effects other than those of $u$ and $d$ quarks are not included. 

One way to determine the lattice spacings is to use 
the $\rho$ meson mass. We refer the reader to  
Ref.~\cite{burkhalter98,lightquark00,lighthadron00}
for the light hadron spectroscopy calculation, from which
we constructed Tables~\ref{tab:lat_physical_dyn} and 
\ref{tab:lat_physical_que}.

Another way is to use $\Upsilon$ level splittings. 
Specifically, we take the spin-averaged $1P-1S$ splitting, which
is considered to be relatively insensitive to
systematic errors.
Our study of the $\Upsilon$ spectroscopy
using the NRQCD action 
is  described in Ref.~\cite{manke00}.

One may think that the most natural scale for the physics of $B$ mesons 
is their level splittings. 
In quenched
studies~\cite{glok_spectrum00,sgo_spectrum00}, it was
found that the scale from the spin-independent $B$ spectrum agrees 
with that from the $\rho$ meson mass. 
We defer a verification in full
QCD to a separate work. 
In this study, we study the scale from the $\rho$ meson mass,
denoted as $a_\rho$ in the followings, and that from the $\Upsilon$ 
$1P-1S$ splitting, denoted as $a_\Upsilon$.
We do not adopt the string tension to fix the scale
because its physical value is not well known.

\begin{figure}[tbp]  
  \begin{center}
    \epsfclipon
    \epsfxsize=0.6\hsize 
    \epsfbox{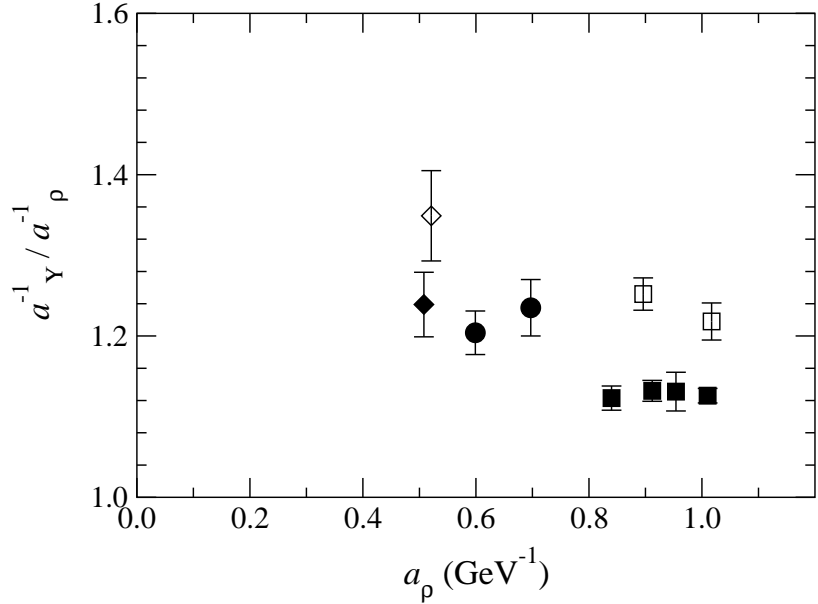}
  \end{center}
  \caption{
    Ratio of inverse lattice spacings from $\Upsilon(1P-1S)$
    and $m_\rho$ for full (filled symbols) and quenched
    (open symbols) QCD.
    Our results for $N_f$ = 2 (partially quenched)
    lattices are denoted by filled squares ($\beta$ = 1.95)
    and by filled circles ($\beta$ = 2.1) 
    \protect\cite{manke00}.
    The open and filled diamonds denote a quenched
    \protect\cite{alikhan98,davies98} and partially quenched 
    \protect\cite{sara99} result respectively, for the
    plaquette gauge action.
    Error bars are purely statistical.
    }
  \label{fig:scaleratio}
\end{figure}

Our results for the Lattice spacings are listed in
Tables~\ref{tab:lat_physical_dyn} and
\ref{tab:lat_physical_que} for unquenched and quenched
lattices respectively.
Note that $a_\rho$ given in this paper differs
slightly from the one presented in Ref.~\cite{manke00},
the latter being calculated in the chiral limit 
where the $(u,d)$ quark mass vanishes rather than at the physical 
point. The ratio of the scales is plotted in Fig.~\ref{fig:scaleratio}
for quenched (open symbols) and full (filled symbols) QCD. 
The ratio becomes closer to unity with inclusion of the dynamical quark,
but the discrepancy still remains significant. 
We note that the discrepancy does not decrease towards the continuum limit.

The light quark mass corresponding to the $u,d$ quarks is determined from
$m_\pi$. 
To determine the strange quark mass, we use either the $K$ meson mass 
or the $\phi$ meson mass. 
The corresponding hopping parameters 
denoted by $K_l$, $K_s(K)$ and $K_s(\phi)$, are given in
Tables~\ref{tab:lat_physical_dyn} and
\ref{tab:lat_physical_que}.

\subsection{$B$ meson masses}
\begin{figure}[tbp]  
  \begin{center}
    \epsfclipon
    \epsfxsize=0.6\hsize \epsfbox{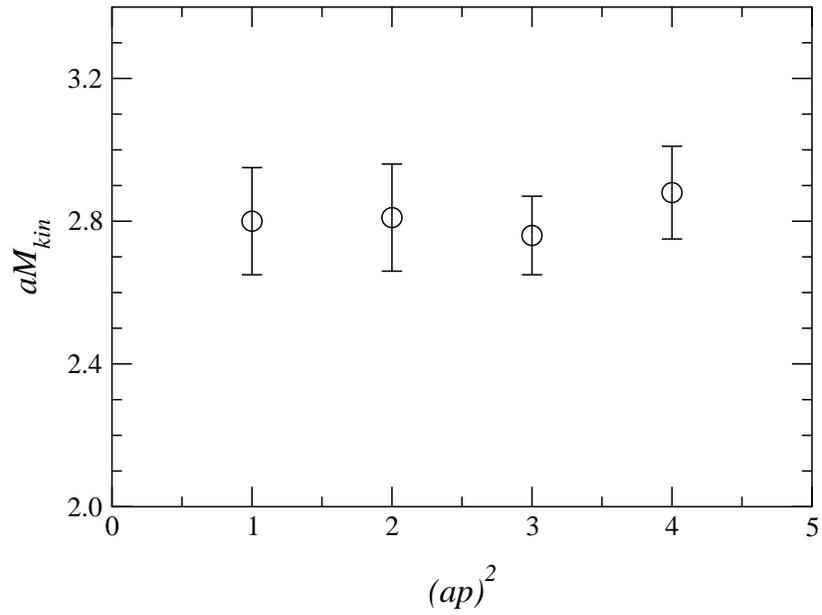}
  \end{center}
  \caption{
    Kinetic mass at $\beta$ = 2.575, extracted from
    Eq.~\protect\ref{eq:Mkin}, in lattice units, as a
    function of the momentum $p^2$ in units of
    $(2\pi/L)^2$. 
    The quark mass parameters are $K_{sea} = 0.1344$ and
    $aM_0 = 2.6$.
    }  
  \label{fig:Mkin2.575}
\end{figure}

In NRQCD, the exponential falloff of the correlator in Euclidean time,
$E_{sim}$, represents the bare binding energy. 
We expect that the nonperturbative mass of 
heavy-light mesons is inferred from the meson dispersion relation.
We use the relativistic form
\begin{equation}
  \Delta E(p^2) \equiv E_{sim}(p^2) - E_{sim}(0)
  = \sqrt{M^2_{kin} + p^2} - M_{kin}
  \label{eq:Mkin}. 
\end{equation}
In practice, we determine this energy difference from a fit
of the ratio of the correlators at $p^2=(2\pi/La)^2$ and
$p^2=0$ to a single exponential. 
The results are given in Table~\ref{tab:Mkin_dyn} for full
QCD and in Table~\ref{tab:Mkin_quenched} for the quenched
case in the Appendix.
We also examine this particular form of the dispersion
relation by comparing the results using momenta larger than
one lattice unit, and find that they agree within errors. 
An example for $M_{kin}$ as a function of $p^2$ is given 
in Fig.~\ref{fig:Mkin2.575} for a quenched lattice at
$\beta$ = 2.575.

The meson masses can also be estimated from $E_{sim}$, through the
perturbative relation
\begin{equation}
  M_{pert} = E_{sim} + \Delta_{pert}
  \equiv E_{sim} + Z_m M_0 - E_0,
\end{equation}
where $Z_m$ is the quark mass renormalization constant, and
$E_0$ is a shift of the zero point of the energy that occurs in
non-relativistic and static theories.
We employ one-loop perturbative values of $Z_m$ and $E_0$
\cite{ishikawa99},  using $\aMSbar$ as defined in
(\ref{eq:alpha}) at the scale $1/a$.
Results for $E_{sim}$ are given in 
Tables~\ref{tab:Esim_dyn} and \ref{tab:Esim_quenched}, 
and those for $M_{pert}$ in
Tables~\ref{tab:Mpert_dyn} and \ref{tab:Mpert_que} in the Appendix.

The statistical errors in $E_{sim}$ are very small. The
error of $M_{pert}$ quoted in these tables is dominated by
the systematic error from higher order radiative
corrections, as estimated by $\aMSbar^2(1/a)$ times the meson
mass.
We find that the one-loop contribution to $\Delta_{pert}$ is
always smaller than our estimate of the two-loop error,
which increases our confidence in the error estimate. 

\begin{figure}[tbp]
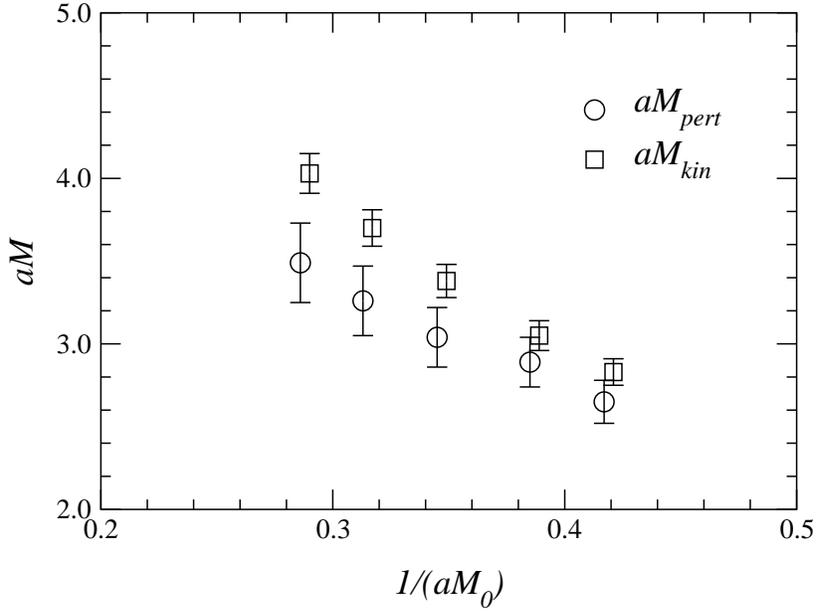
  
  \begin{center}
    \epsfclipon
    \epsfxsize=0.6\hsize 
    \epsfbox{./Mkin_Mpert_beta2.1_1382.eps}
    \\[5mm]
    \epsfxsize=0.6\hsize
    \epsfbox{./Mkin_Mpert_beta2.575.eps}
  \end{center}
  \caption{
    Comparison of kinetic mass (circles) and mass using
    perturbative shifts (squares) in full QCD 
    at $\beta$ = 2.1,  $K_{sea}$ = 0.1382 and $K_{val}$ =
    0.1374 (top panel), 
    and in quenched QCD at $\beta$ = 2.575, $K$ = 0.1344
    (bottom panel).
    } 
  \label{fig:MkinMpert}
\end{figure}

For light valence quark masses around $m_s$, 
$M_{kin}$ and $M_{pert}$ agree within the combined
errors for all configurations except for those for $\beta$ =
2.1, $K_{sea}$ = 0.1357 in full QCD, and for $\beta$ =
2.575 in quenched QCD.
Even for these cases the difference is at most 2 standard
deviations of the statistical error in $M_{kin}$.
In Fig.~\ref{fig:MkinMpert} we show a comparison between
$M_{kin}$ and $M_{pert}$ for full (top panel) and
quenched QCD (bottom panel) at our finest lattice spacing of
$a^{-1}_\rho\approx 1.8$ GeV.
The full QCD data show an agreement which is typical of our data,
while for the quenched data we show the case of the largest discrepancy.

To determine the bare $b$ quark mass, $M_{0b}$, we employ the
kinetic meson mass $M_{kin}$, as it is free from higher order
perturbative errors. The
systematic uncertainty in the choice of the method will be
discussed later. 
We first fit the mass as a linear function of the light
quark mass, 
\begin{equation}
  M = A_q + \frac{B_q}{2}
  \left(\frac{1}{K} - \frac{1}{K_c}\right),
\end{equation}
and extrapolate or interpolate to the physical value $K_l$
and $K_s$ to obtain the heavy-light meson masses $M_l$ and
$M_s$.
The result is then expressed as a function of the heavy quark mass, as 
\begin{equation}
  \label{eq:fit_1/M}
  M_{l,s} = A_Q M_0 + B_Q,
\end{equation}
and $M_{0b}$ is determined by requiring $M_l$ or $M_s$ to
equal the physical meson mass, $M_{B}$ or $M_{B_s}$,
respectively.

\begin{figure}[tbp]
  \begin{center}
    \centerline{\epsfysize=0.4\vsize
      \epsfbox{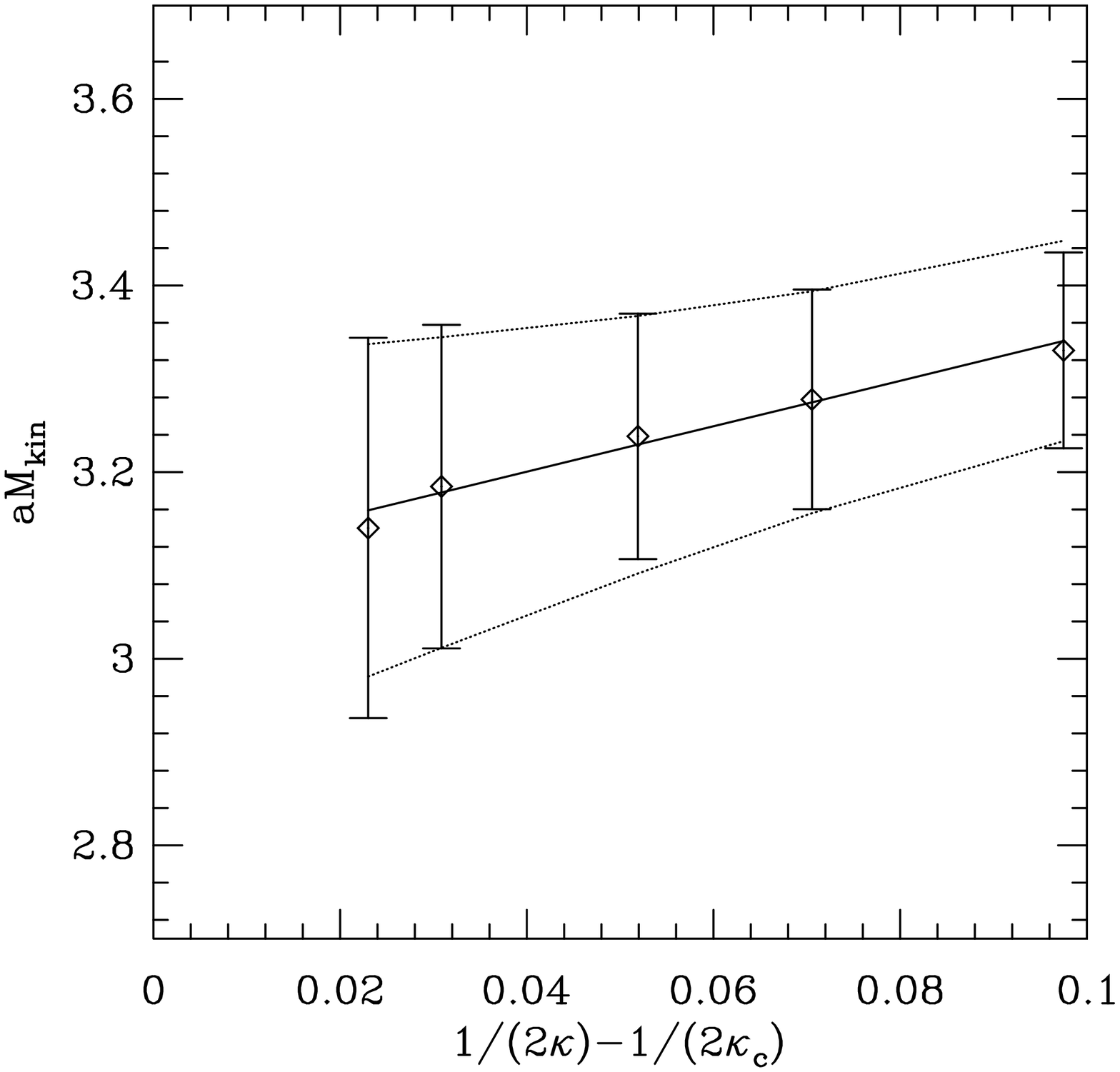} 
      \epsfysize=0.4\vsize
      \epsfbox{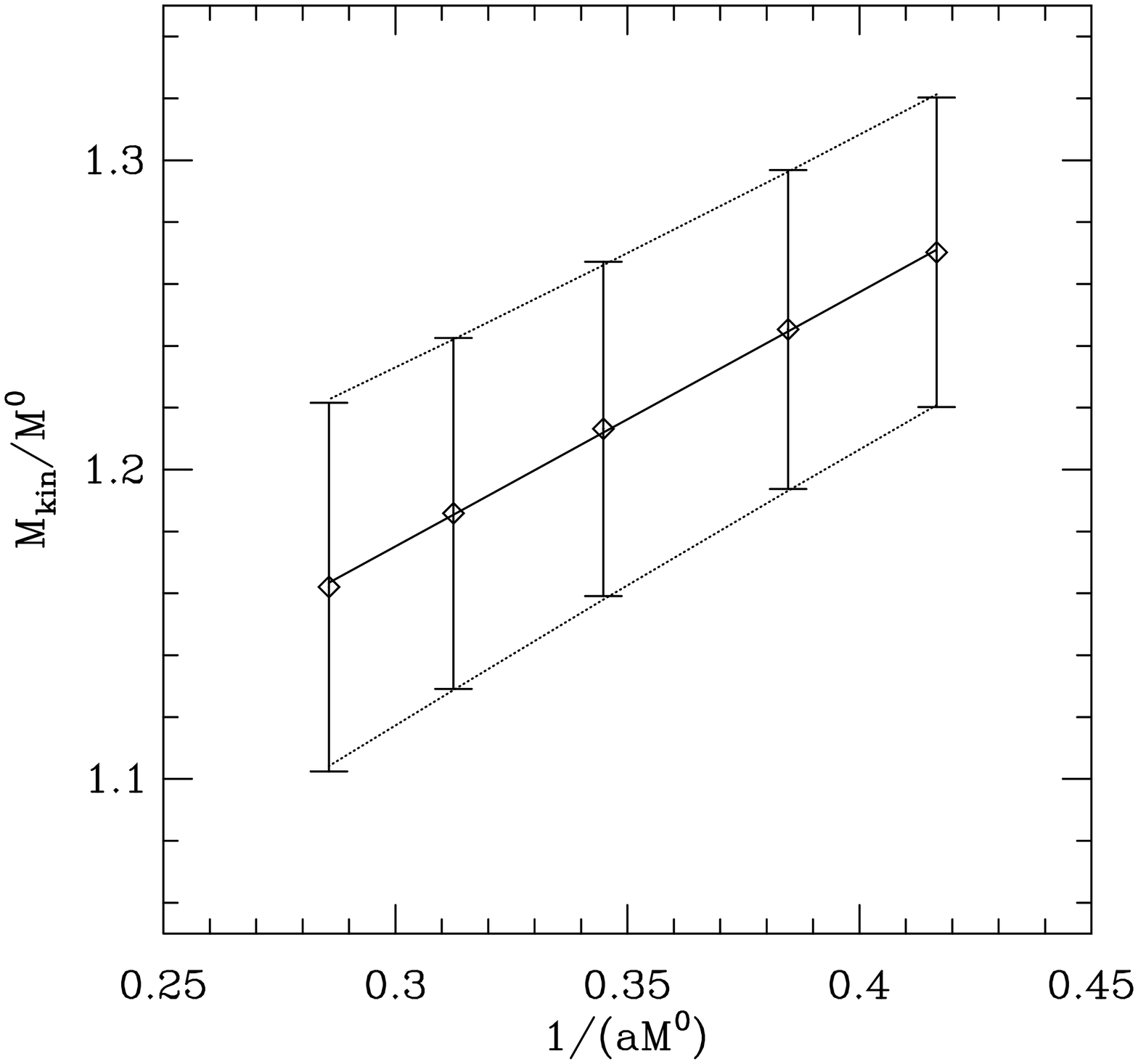} 
      }
  \end{center}
  \caption{
    Kinetic meson mass for the parameter values $\beta$ = 
    2.1, $K_{sea}$ = 0.1374 in full QCD. 
    Left panel shows  a fit of $M_{kin}$ as a function of the
    light quark mass for $aM_0$ = 2.6, which is
    close to $M_{0b}$.  
    Right panel shows a fit in the heavy quark mass
    with the light quark mass interpolated to the strange
    quark mass.
    Solid lines denote the fits and dashed lines the
    error. 
    } 
  \label{fig:MHL_vs_mquark}
\end{figure}

Examples for these fits in the light and heavy quark mass
are given in Fig.~\ref{fig:MHL_vs_mquark}.
On the right panel, a plot of $M_{s}/M_0$ is shown as a
function of $1/M_0$, which is $A_Q + B_Q/M_0$.
Results using the $B$ meson agree with those from $B_s$
allowing for larger errors. 
We use the $B_s$ rather than the $B$ meson to calculate the
central values of $M_{0b}$ to avoid the larger statistical
and possible systematic errors from the extrapolation to
$K_l$. 
The difference between the use of the $K$ and
$\phi$ mesons to fix the strange quark mass is negligible
compared to other errors in heavy-light meson mass. 
We take the central value from the $K$ meson.
The numerical results for $M_{0b}$ are listed in
Table~\ref{tab:Mb0_PQ} (full) and in Table \ref{tab:Mb0_que} 
(quenched).

\subsection{Decay constants}
The decay constants are calculated from the results of the fit
(\ref{eq:C_SS}) and (\ref{eq:C_SL}) according to
\begin{equation}
  \label{eq:fsqrtM}
  a^{3/2} (f\sqrt{M})^{(i)} = 
  \frac{1}{\sqrt{M}} \langle 0|J_L^{(i)}|P(\vec{0})\rangle
  =
  \sqrt{2} Z_L^{(i)} \sqrt{1-\frac{3K}{4K_c}},
\end{equation}
where $|P(\vec{0})\rangle$ denotes a pseudoscalar
heavy-light meson state of mass $M$ at rest.
The normalization factor $\sqrt{1-\frac{3K}{4K_c}}$ for the
light quark field in (\ref{eq:fsqrtM}) is motivated by the
meanfield improvement of the perturbative renormalization
factor \cite{lepmac93}, with which the one-loop coefficient
$\rho_A^{(0)}$ in (\ref{eq:matching}) is modified.
The lattice results for the matrix elements
$(f\sqrt{M})^{(0)}$ and $(f\sqrt{M})^{(1)}$ are listed in
Tables~\ref{tab:frootm_beta1.95}--\ref{tab:frootm_beta2.1}
for full QCD and in
Tables~\ref{tab:frootm_beta2.187}--\ref{tab:frootm_beta2.575}
for quenched QCD.
As mentioned before in Sec.~\ref{sec:formalism:currents}, 
$(f\sqrt{M})^{(2)} = (f\sqrt{M})^{(1)}$ at zero momentum.

\begin{figure}[tbp]  
  \begin{center}
    \epsfclipon
    \epsfxsize=0.6\hsize
    \epsfbox{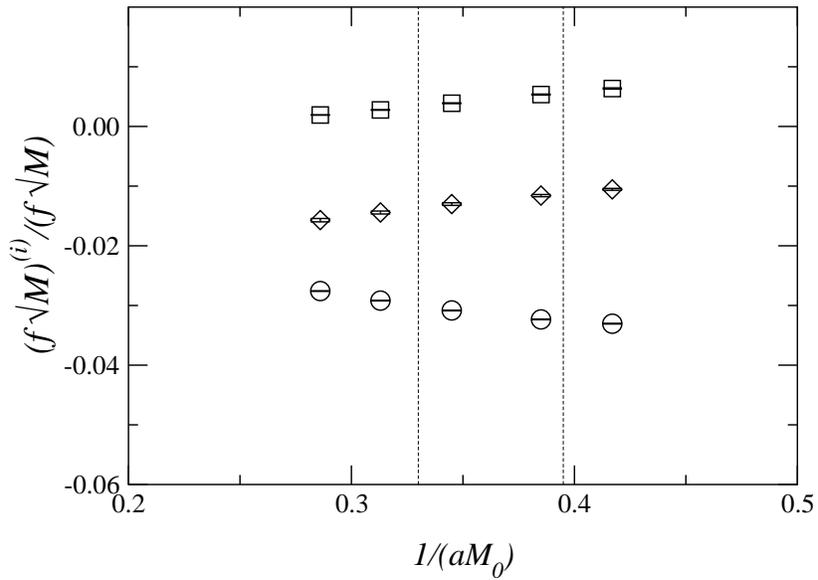}
  \end{center}
  \caption{
    Relative size of the one-loop corrections to the current
    matrix elements in full QCD. 
    Circles denote 
    $\alpha_s\rho_0(f\sqrt{M})^{(0)}/(f\sqrt{M})$, 
    squares stand for
    $\alpha_s\rho_1(f\sqrt{M})^{(1)}/(f\sqrt{M})$, 
    and diamonds give
    $\alpha_s\rho_2(f\sqrt{M})^{(2)}/(f\sqrt{M})$. 
    The two vertical lines is a band to indicate the
    position of the bare $b$ quark mass. 
    Parameter values are $\beta = 2.1$, 
    $K_{sea} = K_{val} = 0.1382$. 
    }
  \label{fig:oneloop}
\end{figure}

The physical decay constant is then obtained, following (\ref{eq:matching}), 
by 
\begin{equation}
  \label{eq:mixing}
  f\sqrt{M} = 
  (1+\alpha_s \rho_A^{(0)}) (f\sqrt{M})^{(0)} +
  (1+\alpha_s \rho_A^{(1)}) (f\sqrt{M})^{(1)} +
  \alpha_s \rho_A^{(2)} (f\sqrt{M})^{(2)}.
\end{equation}
The one-loop corrections of the different currents
contribute with different signs and partially cancel each
other. 
As shown in Fig.~\ref{fig:oneloop} as an example for full
QCD, the relative contribution of each of these corrections
to the whole decay matrix element $f\sqrt{M}$ is small. 
The largest is the correction to $(f\sqrt{M})^{(0)}$,
being of the order of a few percent. 
The overall one-loop correction is $3-6\%$ for the dynamical
case, and $3-5\%$ for quenched.
The one-loop correction for the RG-improved gauge action is
thus smaller than for the plaquette gauge action for which
the correction amounts to roughly $10\%$ 
(see e.g.~\cite{alikhan98,JLQCD00}). 

\subsection{Analysis of heavy and light quark mass dependence}

\begin{figure}[tbp]
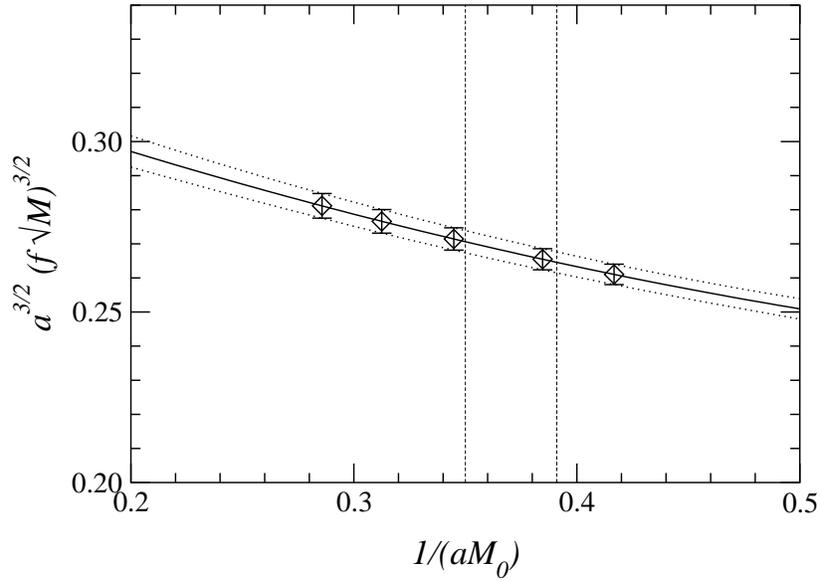

  \begin{center}
    \epsfclipon
    \epsfxsize=0.6\hsize
    \epsfbox{./f0rootm_vs_mq_beta2.1_1374_2.6.eps}
    \\[5mm]
    \epsfxsize=0.6\hsize
    \epsfbox{./frootm_vs_1overM_beta2.1_1374.eps} 
  \end{center}
  \caption{
    Decay matrix elements for the parameter values $\beta$ = 
    2.1, $K_{sea}$ = 0.1374. 
    Top panel shows a fit of $(f\protect\sqrt{M})^{(0)}$
    as a function of the light quark mass for 
    $aM_0$ = 2.6, which is close to $M_{0b}$.
    Bottom panel shows a fit of $(f\protect\sqrt{M})$
    in the heavy quark mass with the light quark mass
    interpolated to the strange quark mass. 
    Solid curves denote the fits and dotted lines the
    error.
    The vertical lines in the figure on the bottom give the 
    error bounds of $aM_{0b}$.
    } 
  \label{fig:frootm_vs_mquark}
\end{figure}

We find that the decay matrix elements
$a^{3/2}(f\sqrt{M})^{(i)}$ are well described by a linear
function in the light quark mass $1/(2K)-1/(2K_c)$ as shown
in Fig.~\ref{fig:frootm_vs_mquark} (top panel).
Using the linear fit, we interpolate the data to $K_s$ or
extrapolate to $K_l$.
Then, we make a quadratic fit in $1/M_0$ and interpolate
to the $b$ quark mass, $1/M_{0b}$.
Figure~\ref{fig:frootm_vs_mquark} (bottom) shows this fit
for full QCD on the finest lattice.

For quenched QCD this completes the analysis of the decay
constant. 
We give the renormalized decay constants in
Table~\ref{tab:resquenched}.

\begin{figure}[tbp] 
  \begin{center}
    \epsfclipon
    \epsfxsize=0.6\hsize
    \epsfbox{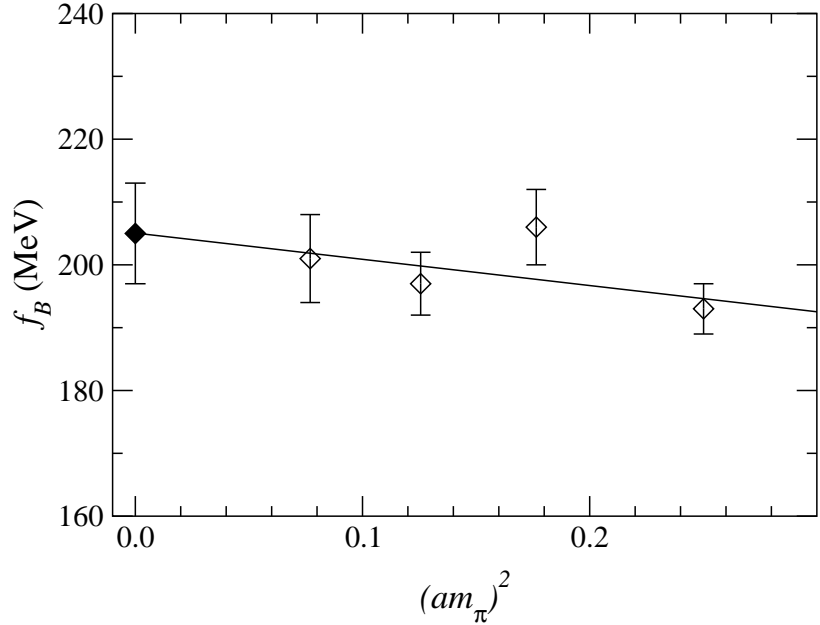} 
  \end{center}
  \caption{
    Decay constant at $\beta = 2.1$ as a function of the sea
    quark mass. 
    Open symbols denote the partially quenched results, the
    solid line the fit in $(am_\pi)^2$, and the filled
    symbol the value in the chiral limit of the sea quark
    mass.
    } 
  \label{fig:fB_vs_msea}
\end{figure}

For full QCD we carry out this procedure separately
for each sea quark mass $m_{sea}$, using partially quenched
values for the lattice spacing and $K_c$, $K_l$ and $K_s$. 
The resulting renormalized decay
constants are given in Table~\ref{tab:respartquenched}.
From these data we calculate the physical decay
constants by extrapolating the sea quark mass to the $u,d$
quark mass.
We use a linear fit in $(am_\pi)^2$, where $m_\pi$ is the
pion mass of the sea quark. 
The sea quark mass dependence of $f_B$, $f_{B_s}$,
and $f_{B_s}/f_B$, is very mild. 
On finer lattices, there is a slight upwards shift of the
decay constants as the sea quark mass is decreased. 
In Fig.~\ref{fig:fB_vs_msea}, this is demonstrated for
$f_B$ at $\beta$ = 2.1. 
The final results for full QCD are given in
Table~\ref{tab:resdyn}.

We note that the decay constants we obtained for finite lattice
spacings agree very well with those from the Fermilab
formalism \cite{hugh} for both quenched and full calculations.

We add comments on possible systematic errors
from various steps of the analysis procedure: 
(i) The bare $b$ quark mass using the $\rho$
mass to set the scale is slightly higher than 
that from the $\Upsilon$, but the two agree within 
statistical errors. 
(ii) The statistical error on the mass is included within
the bootstrap procedure. 
There is a good agreement between results from $M_{kin}$ and
$M_{pert}$, and if the perturbative error is included in the
determination using $M_{pert}$, their errors are very
similar. 
We take results obtained with the kinetic masses for our
central values.
(iii) An alternative method to
represent the heavy quark dependence of $f^{(i)}\sqrt{M}$ is to
employ the heavy-light meson mass instead of the heavy quark
mass.
This circumvents the determination of $M_{0b}$. 
The results agree with those using $M_{0b}$  within the
statistical errors, which are similar in magnitude in both methods. 
We quote the numbers from the procedure using $M_{0b}$ as
our central values.

\section{Estimate of continuum results}
\label{sec:continuum}

While discretization errors decrease as the lattice spacing
becomes smaller, the $1/M$ operators introduce radiative
corrections that go as powers of $1/a$ in NRQCD.
Thus NRQCD breaks down in the limit of vanishing lattice
spacing, and we cannot perform a continuum extrapolation
as a matter of principle.
The aim is rather to find a range of lattice spacings where the result
is approximately independent of the cutoff with small systematic
errors.

\begin{figure}[p]
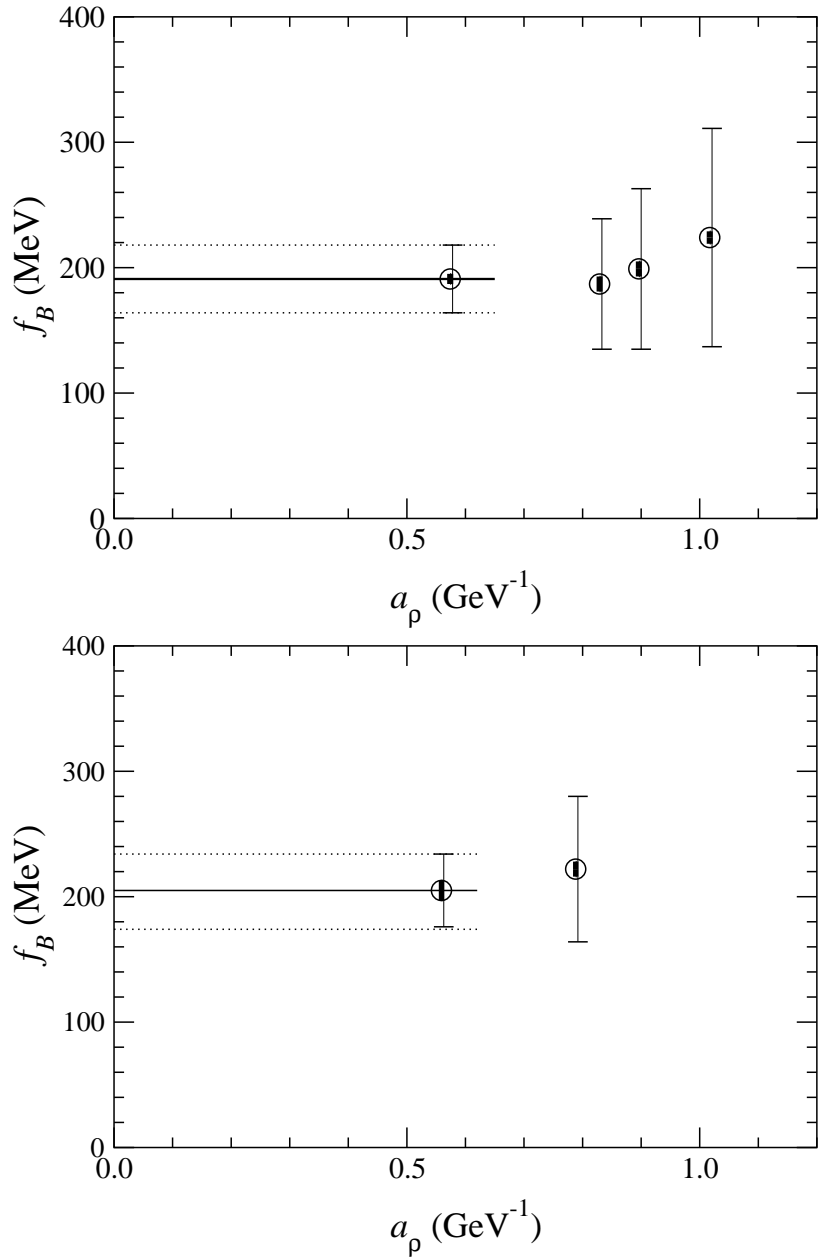
  
  \begin{center}
    \epsfclipon
    \epsfxsize=0.6\hsize
    \epsfbox{./fb_nrqcd_quenched_vs_arho.eps}
    \epsfxsize=0.6\hsize
    \epsfbox{./fb_nrqcd_dynamical_vs_arho.eps}
  \end{center}
  \caption{
    $f_B$ as a function of the lattice spacing in quenched QCD (top)
    and full QCD (bottom). 
    Thick error bars  denote statistical, thin error
    bars, systematic errors.
    The solid horizontal line shows the final estimate for
    $f_B$ taken from the finest lattice, the dashed horizontal lines, the
    error     (statistical and systematic added in quadrature) on 
    this estimate.
    }
  \label{fig:fbscaling}
\end{figure}

\begin{figure}[tbp]  
  \begin{center}
    \epsfclipon
    \epsfxsize=0.6\hsize
    \epsfbox{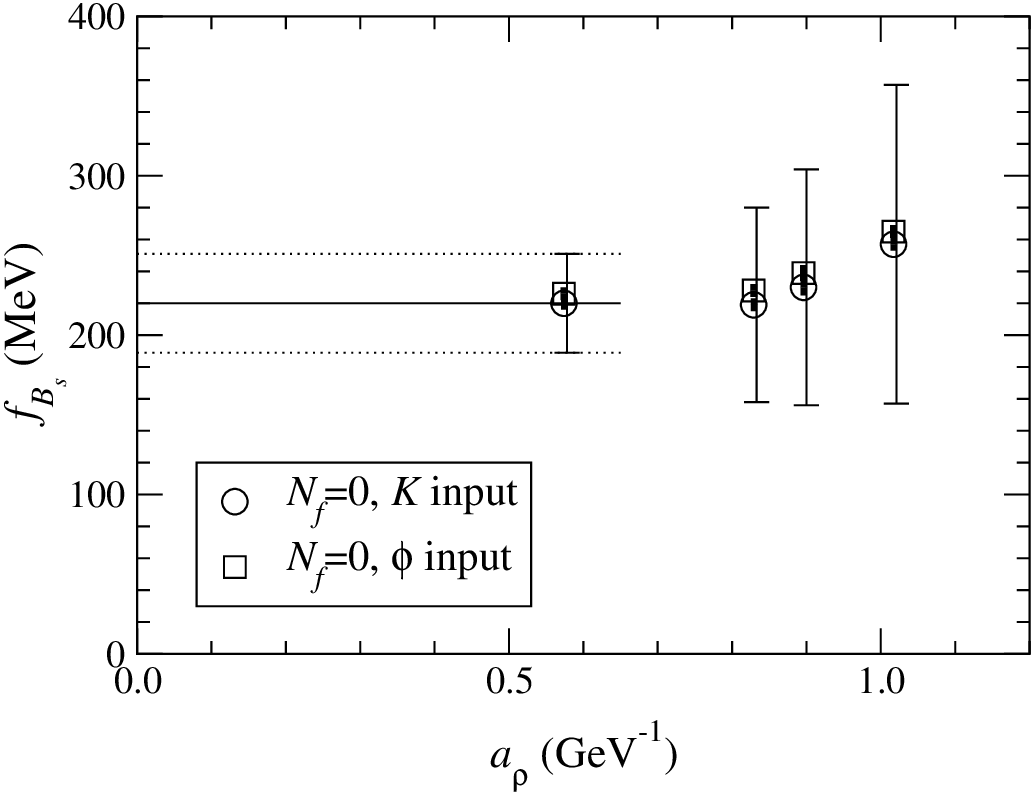} 
    \epsfxsize=0.6\hsize
    \epsfbox{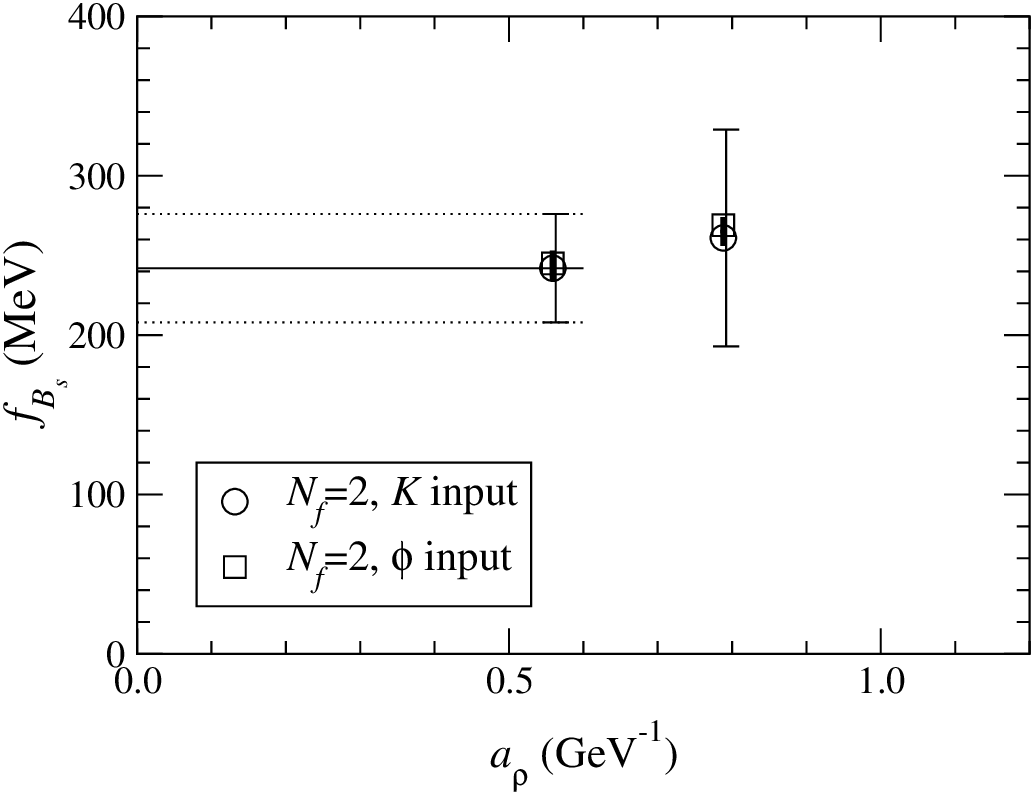}
  \end{center}
  \caption{
    $f_{B_s}$ as a function of the lattice spacing in quenched QCD (top)
    and full QCD (bottom). Thick error
    bars denote statistical errors, thin ones, systematic errors.
    The solid horizontal line shows the final estimate for
    $f_{B_s}$ taken from the result on the finest lattice, the dashed
    horizontal lines, the error on   this estimate.
    } 
  \label{fig:fbsscaling}
\end{figure}

We have plotted the resulting $f_B$ and $f_{B_s}$ at finite lattice spacing
in  Fig.~\ref{fig:fbscaling} and \ref{fig:fbsscaling} for both quenched
and full QCD calculations, where statistical errors are shown with  
thick bar symbols. We see the data for quenched QCD (top panels)  
exhibiting a signature for a plateau for $a_\rho\leq 0.8$ GeV$^{-1}$. 
The data for full QCD
calculation however show a sizable drop from $a_\rho=0.79$ to 0.56 GeV$^{-1}$.
Nevertheless, it is perhaps possible to estimate the continuum value if
we can properly estimate the discretization error at each lattice
spacing, and we consider this problem in the following.

The leading systematic errors due to finite lattice spacings 
are $O(\alpha_s a\Lambda_{QCD})$ and $O(a^2\Lambda_{QCD}^2)$ 
with the clover action for light valence quarks. 
In the NRQCD heavy quark action, the tree level matching of
the chromomagnetic coupling $c_B$ leads to the error of
$O(\alpha_s\Lambda_{QCD}/M_b)$. 
We also expect the discretization error of
$O(\Lambda_{QCD}a/M_b)$ from the NRQCD action.
Since the matching calculation of the axial current has been
made in one-loop perturbation theory, we expect an
$O(\alpha_s^2)$ radiative correction. 
In addition, there is an $O(\alpha_s^2/(aM_b))$ term, which
is formally subleading compared to the $O(\alpha_s^2)$ term,
but still gives one of the leading $1/a$ contributions. 
An $O(\Lambda_{QCD}^2/M_b^2)$ contribution comes from the
neglected $1/M^2$ corrections to the NRQCD Lagrangian.
In Table~\ref{tab:syst_errors_600} we list an estimate
of the systematic errors for $f_B$ and $f_{B_s}$ on our
lattices.
We use $\aMSbar(1/a)$ for the coupling constant $\alpha_s$.
For the typical momentum scale $\Lambda_{QCD}$, we naively
expect that its order is around 200-300~MeV.
In this analysis, however, we take a value which is larger
by a factor two and use $\Lambda_{QCD}$ = 600~MeV, in order
to make the estimate more conservative.  
The overall error is calculated by summing the individual
contributions in quadrature.

In Figs.~\ref{fig:fbscaling} and \ref{fig:fbsscaling} 
the discretization errors thus estimated are shown with  
thin error bars. Looking first at figures for the quenched case
(top panels) we see that the ranges of error bars
overlap among all data points, but also that the value of plateau
is within the ranges of error bars.
Therefore, we consider that the continuum value of the quenched decay
constant is located within the estimated errors. 
We take the data at the weakest coupling, with
which the error estimate becomes minimum, 
\begin{eqnarray}
  f_B^{N_f=0}     &=& 191 \pm 4 \pm 27\, {\rm MeV},\\
  f_{B_s}^{N_f=0} &=& 220 \pm 4 \pm 31\, {\rm MeV},
\end{eqnarray} 
as our estimate for the continuum value, where the first
error is statistical, and the second is uncertainty
associated with the discretization. 
These values are shown by horizontal lines in the figure. 
The errors here do not include systematic errors from
the uncertainties of the strange quark mass (3\%) and the 
lattice scale which exceeds 30\% (see Fig.~\ref{fig:scaleratio}). 
The error due to strange quark mass is ascribed 
to both lattice artifacts and quenching effects. 
As seen in Table~\ref{tab:resquenched} this uncertainty
decreases towards weaker couplings. 
On the other hand, the uncertainty due to lattice 
scale does not diminish from strong to weaker couplings.

\begin{figure}[tbp]  
  \begin{center}
    \epsfclipon
     \epsfxsize=0.6\hsize 
     \epsfbox{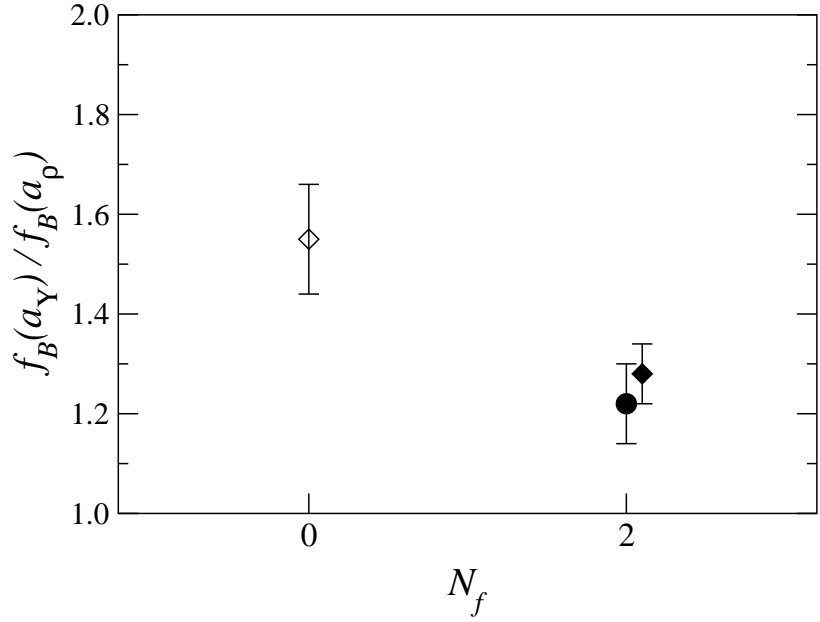}
  \end{center}
  \caption{
    Ratio of decay constants determined with $a_\Upsilon$
    and $a_\rho$ for quenched ($N_f$ = 0) and partially
    quenched ($N_f$ = 2) QCD.
    Filled circle is our results, while filled diamond is
    from \protect\cite{sara99}.
    The quenched data (open diamond) has been obtained reanalyzing
    the results from \protect\cite{alikhan98} using lattice spacings from the
    $\Upsilon$. 
    }
  \label{fig:fbvsnf}
\end{figure}

For the full QCD case, we could not see a plateau within the 
statistical errors in Figs.~\ref{fig:fbscaling} and \ref{fig:fbsscaling}.
From the same reasoning as in the quenched case, however, we expect that the 
continuum value is within the error range when we include 
systematic errors. 
Indeed, the final error bars of the two data points largely
overlap in Figs.~\ref{fig:fbscaling} and \ref{fig:fbsscaling}. 
Taking the data with a smaller error bar 
({\it i.e.} at the weaker coupling), we have
\begin{eqnarray}
  f_B^{N_f=2}     &=& 204 \pm 8 \pm 29\, {\rm MeV},\\
  f_{B_s}^{N_f=2} &=& 242 \pm 9 \pm 34\, {\rm MeV},
\end{eqnarray}
as our `final' estimate. 
We take these value as still provisional, since
a plateau is not identified within the statistical errors. 
In full QCD, the uncertainty from the strange quark mass is about 
1\%. 
The uncertainty from the lattice scale is 22\% in full QCD, 
which is smaller than that in the quenched case, but is still substantial
(see Fig.~\ref{fig:fbvsnf}). 
To be conservative we add this error to the final estimate of errors for 
the decay constants.

Figure \ref{fig:fbsoverfbscaling} shows a similar analysis 
for the ratio $f_{B_s}/f_B$.
This ratio is rather insensitive to the perturbative
corrections and the heavy quark action. 
The dominant errors come from the light quark action,
i.e. $O(\alpha_s a \Lambda_{QCD})$ and
$O(a^2\Lambda_{QCD}^2)$.
We estimate the systematic error of $f_{B_s}/f_B-1$ by
quadrature.
We find the systematic error to be smallest again at the
finest lattice spacing, whose results are therefore taken as
our final estimates. We obtain

\begin{eqnarray}
  \left(\frac{f_{B_s}}{f_B}\right)^{N_f=0}  & = & 1.150\pm 0.009\pm 0.020, \\
  \left(\frac{f_{B_s}}{f_B}\right)^{N_f=2}  & = & 1.179\pm 0.018\pm 0.023.
\end{eqnarray}

To study the effect of dynamical sea quarks, 
we take the ratio of unquenched to the quenched results.
We expect that most systematic errors cancel, and
we obtain,
\begin{eqnarray}
  \frac{f_B^{N_f=2}}{f_B^{N_f=0}} 
  &=& 1.07(5); \\
  \frac{f_{B_s}^{N_f=2}}{f_{B_s}^{N_f=0}} 
  &=& 1.10(5); \\
  \frac{\left(f_{B_s}/f_B\right)^{N_f=2}}{\left(f_{B_s}/f_B\right)^{N_f=0}} 
  &=& 1.03(2).
\end{eqnarray}
where only the statistical errors are retained.
We observe that the inclusion of dynamical sea quarks increases 
the decay constants, as noted in 
\cite{sara99,MILC99}. The effects are 1.5 sigma
for $f_B$, and 2 sigma for $f_{B_s}$.

\section{Comparison with other studies}
\label{sec:comparison}
We first note that the decay constants we obtained with the NRQCD
formalism agree very well with those obtained using the Fermilab
formalism; $f_B=188\pm3\pm26$ MeV and $f_{B_s}=220\pm2\pm31$ MeV
from a quenched calculation, and   
$f_B=208\pm10\pm29$ MeV and $f_{B_s}=250\pm10\pm35$ MeV from a full
QCD calculation \cite{hugh}. This justifies the 
validity of the two formalisms within the statistic and systematic 
errors.

\begin{figure}[tbp]  
  \begin{center}
    \epsfclipon
    \epsfxsize=0.6\hsize \epsfbox{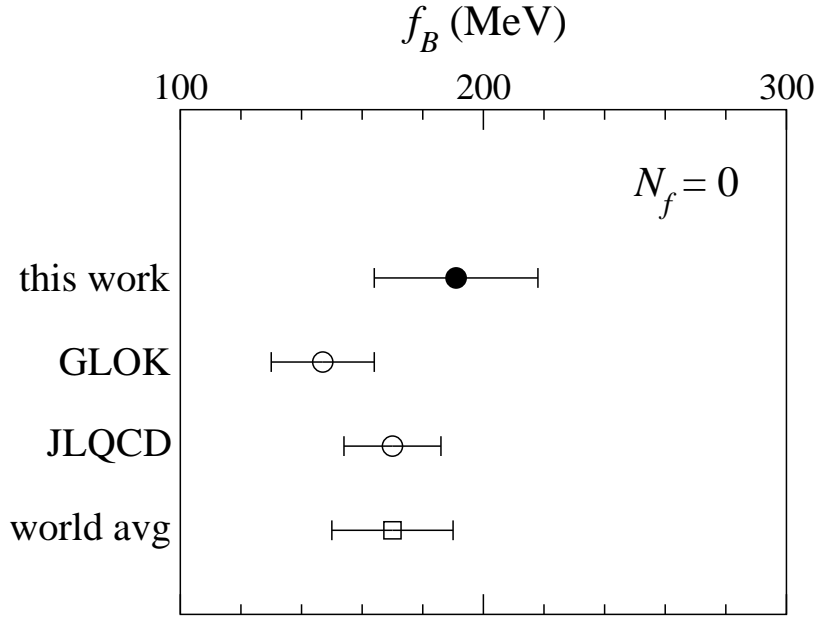}
  \end{center}
  \caption{
    Comparison of quenched results for $f_B$ from NRQCD. 
    The filled circle denotes the quenched result quoted in
    this paper, while two open circles show results from
    other recent NRQCD studies~\protect\cite{alikhan98} and
    \protect\cite{JLQCD00}.
    The open square stands for the quenched world average
    quoted in~\protect\cite{hashimoto99}. 
    Errors include statistical and systematic errors
    combined in quadrature.
    }
  \label{fig:fbquenched_overview}
\end{figure}

We now compare our quenched calculation to the previous NRQCD results,
\cite{alikhan98} and \cite{JLQCD00}, and to a quenched
world average \cite{hashimoto99} in
Fig.~\ref{fig:fbquenched_overview}.   
In Ref.~\cite{alikhan98}, marked as GLOK in 
Fig.~\ref{fig:fbquenched_overview}, the
clover light quark action is used with a tree-level meanfield-improved
clover coefficient, and the NRQCD action includes
$1/M^2$ corrections at the tree-level. 
The lattice spacing, fixed from $m_\rho$, is $a^{-1}\simeq2$ GeV. 
The calculation of the JLQCD Collaboration \cite{JLQCD00}
employed the clover light quark action with a
meanfield-improved one-loop clover coefficient, and heavy
quarks with NRQCD corrected through $O(1/M)$. 
The scaling of the decay constant is tested at three lattice
spacings, and the central value is extracted from runs at a
lattice spacing $a^{-1} \simeq 1.6 $ GeV, with the string
tension used to set the scale.
The scaling test is also made for $f_{B_s}$ in
Ref.~\cite{GLOK_scaling} in the range $a^{-1}$=1.1 to 2.6~GeV. 
For the world average, we use the value quoted in
Ref.~\cite{hashimoto99} that includes a variety of quark
actions. 
Our value (labeled as ``this work'') is slightly higher than those of
JLQCD and the world average by 1.5 sigma. Our value, however, disagrees
with the GLOK result by 2.5-3 sigma.

\begin{figure}[tbp]  
  \begin{center}
    \epsfclipon
    \epsfxsize=0.6\hsize \epsfbox{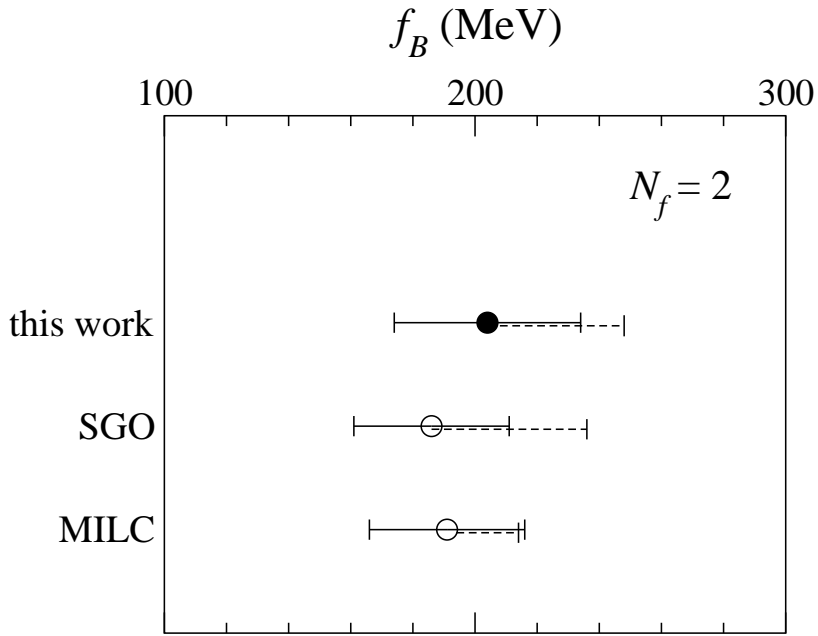}
  \end{center}
  \caption{Comparison of  results for $f_B$ in $N_f = 2$ QCD. The filled
    circle denotes the result quoted in this paper for $N_f = 2$, the two open
    circles stand for the results from~\protect\cite{sara99}
    and~\protect\cite{MILC99}. 
    Solid error bars include statistical and systematic errors combined in
    quadrature. The uncertainty from setting the lattice spacing from
    light physics quantities or $\Upsilon$ spectroscopy
    is shown separately in dotted lines.}   
  \label{fig:fbdynamical_overview}
\end{figure}

In Fig.~\ref{fig:fbdynamical_overview}, we compare our
result from two-flavor QCD with other dynamical calculations 
that use the plaquette gauge action. 
Ref.~\cite{sara99} denoted as SGO uses the NRQCD action corrected
through $O(1/M)$ for the heavy quark and a
tree-level clover action for the light valence quark. 
The calculation used a set of dynamical gauge configurations with
staggered quarks at a mass around the strange
quark mass, but on a somewhat small physical volume 
($L_s \sim 1.6$ fm).
The lattice spacing ($a^{-1} \simeq 2$ GeV) is fixed with
$m_\rho$.   
The result from MILC \cite{MILC99} is an ongoing study using
two flavors of staggered sea quarks, and both Wilson and
(fat-link) clover action for valence quarks. 
The lattice spacings are set using $f_\pi$, and the central
value is extracted from the continuum extrapolation of the
Wilson results.
Our result is again slightly higher, but agrees with the others within
one sigma error. 

\section{Conclusions}
\label{sec:conclusions}
We have calculated the $B$ meson decay
constants in two-flavor full QCD using the $O(1/M)$ NRQCD action,
paying attention to the
sea quark mass dependence and extrapolation to the chiral
limit of the sea quark as well as estimates of discretization
errors.  We have used improved actions for
both quarks and gluons to minimize the discretization error
at a modest lattice spacing. 
We also have performed quenched simulations at matching lattice 
spacings using the same actions, 
to study the effect of dynamical sea quarks.

We have confirmed a plateau in the plot of the decay constants as
a function of lattice spacing for the quenched calculation, but 
could not see a plateau in the full QCD calculation, within the 
statistical errors.
We have estimated the decay constant in the continuum by evaluating
systematic errors from discretization at each lattice spacing.

We have found that the NRQCD action and the relativistic Fermilab 
formalism give consistent estimates for the $B$ meson decay constants. 
Our values of $f_B$ and $f_{B_s}$ are slightly higher than those 
from previous studies, but the disagreement is
at most at 1 sigma level. 

We have confirmed that the sea
quark effect makes the decay constants larger by $\approx$10\%,
which is about 1.5--2 sigma effects in our statistics.  
The systematic error due to the uncertainty of the strange quark mass 
(whether it is determined from $K$ or $\phi$) is reduced to a negligible 
level by the introduction of two flavors of dynamical sea quarks. 
Comparing the results using the lattice scales from the 
$\rho$ meson mass and the $\Upsilon$ mass splitting, 
we find that the uncertainty from the lattice scale is also smaller 
in full QCD, but remains to give a substantial error
to the decay constants.  

Our final result for the decay constant is
summarized as, 
\begin{eqnarray}
f_B^{N_f=2}     & = & 204\pm8\pm29{+44\atop -0} {\rm MeV}; \\
f_{B_s}^{N_f=2} & = & 242\pm9\pm34{+38\atop -0} {\rm MeV}; \\
\frac{f_{B_s}^{N_f=2}}{f_B^{N_f=2}}  & = & 1.179\pm18\pm23,  
\end{eqnarray}
where the central values are taken from those 
using the lattice scale from the $\rho$ meson mass, because 
the $B$ physics will be governed by the soft hadron phenomena, 
and the uncertainty from the lattice scale is shown by the third error.
This uncertainty cancels out in the ratio $f_{B_s}/f_B$. 
We should take these values as provisional until a 
plateau is actually confirmed with a future dynamical calculation.

\section*{Acknowledgments}

The calculations were performed on the parallel computer CP-PACS at
the Center for Computational Physics, University of
Tsukuba. 
This work is supported in part by the Grants-in-Aid of
Ministry of Education (Nos. 09304029, 10640246, 10640248,
10740107, 11640250, 11640294, 11740162).  
TM and AAK are supported by the JSPS Research for the Future
Program (Project No. JSPS-RFTF 97P01102).
SE, KN and HPS are JSPS Research Fellows. 

\newpage


\begin{table}
\begin{center}
\begin{tabular}{c|cccc}
\multicolumn{5}{c}{$\beta$ = 1.95 on a $16^3\times 32$ lattice}\\
\hline
$K_{sea}$ & 0.1375 & 0.1390 & 0.1400 & 0.1410 \\
\hline
\# config & 648 & 505 & 670 & 397  \\
\hline
$K_{val}$ & 0.1375 & 0.1375 & 0.1375 & 0.1375 \\
          & 0.1390 & 0.1390 & 0.1390 & 0.1390 \\
          & 0.1400 & 0.1400 & 0.1400 & 0.1400 \\
          & 0.1410 & 0.1410 & 0.1410 & 0.1410 \\
          & 0.1415 & 0.1415 & 0.1415 & 0.1415 \\
\hline
$(aM_0,n)$ & (2.4,2)& (2.4,2)& (2.4,2)& (2.4,2) \\
           & (2.9,2)& (2.9,2)& (2.9,2)& (2.9,2) \\
           & (3.4,2)& (3.4,2)& (3.4,2)& (3.4,2) \\
           & (4.0,2)& (4.0,2)& (4.0,2)& (4.0,2) \\
           & (4.8,2)& (4.8,2)& (4.8,2)& (4.8,2) \\
\hline
\hline
\multicolumn{5}{c}{$\beta$ = 2.1 on a $24^3\times 48$ lattice}\\
\hline
$K_{sea}$ & 0.1357 & 0.1367 & 0.1374 & 0.1382 \\
\hline
\# config &  400 & 403 & 379 & 420 \\
\hline
$K_{val}$ & 0.1357 & 0.1357 & 0.1357 & 0.1357 \\
          & 0.1364 & 0.1364 & 0.1364 & 0.1364 \\
          & 0.1374 & 0.1374 & 0.1374 & 0.1374 \\
          & 0.1382 & 0.1382 & 0.1382 & 0.1382 \\
          & 0.1385 & 0.1385 & 0.1385 & 0.1385 \\
\hline
$(aM_0,n)$ & (2.4,2)& (2.4,2)& (2.4,2)& (2.4,2) \\
           & (2.6,2)&(2.6,2)&(2.6,2)&(2.6,2)\\ 
           & (2.9,2)&(2.9,2)&(2.9,2)&(2.9,2)\\ 
           & (3.2,2)&(3.2,2)&(3.2,2)&(3.2,2)\\ 
           & (3.5,2)&(3.5,2)&(3.5,2)&(3.5,2)\\
\end{tabular}
\end{center}
\caption{Full QCD run parameters.}
\label{tab:para_dyn}
\end{table}
\begin{table}
\begin{center}
\begin{tabular}{c|cccc}
$\beta$ & 2.187 & 2.281 & 2.334 & 2.575 \\
\hline
lattice & $16^3\times 32$ & $16^3\times 32$ & $16^3\times 32$ 
& $24^3\times 48$ \\
\hline
\# config & 195  & 200 & 200 & 200  \\
\hline
$K_{val}$ & 0.1351 & 0.1343 & 0.1337 & 0.1329 \\ 
          & 0.1365 & 0.1357 & 0.1349 & 0.1337 \\
          & 0.1375 & 0.1367 & 0.1358 & 0.1344 \\
          & 0.1385 & 0.1377 & 0.1368 & 0.1351 \\
          & 0.1390 & 0.1383 & 0.1374 & 0.1353 \\
\hline
$(aM_0,n)$ & (2.4,2)& (2.4,2)& (2.4,2)& (2.4,2) \\
           & (2.9,2)& (2.9,2)& (2.9,2)& (2.6,2) \\
           & (3.4,2)& (3.4,2)& (3.4,2)& (2.9,2) \\
           & (4.0,2)& (4.0,2)& (4.0,2)& (3.2,2) \\
           & (4.8,2)& (4.8,2)& (4.8,2)& (3.5,2) \\
\end{tabular}
\end{center}
\caption{Quenched run parameters.}
\label{tab:para_que}
\end{table}
\begin{table}[t]
\begin{center}
\begin{tabular}{c|cccccc|ccl}
&\multicolumn{6}{l|}{Full QCD simulations}&
 \multicolumn{3}{l}{Quenched simulations}\\
\hline
lattice & $\beta$ & $c_{SW}$ & 
$K_{\rm sea}$ & \#traj. & $m_\pi/m_\rho$ & $a_\sigma$~[fm]&
$\beta$ & $c_{SW}$ & $a_\sigma$~[fm]\\
\hline
$12^3{\times}24$ & 1.80 & 1.60 &
        0.1409   & 6250 &   0.806(1) & 0.289(3)&&& \\  
&&&     0.1430   & 5000 &   0.753(1) & 0.280(4)&&& \\
&&&     0.1445   & 7000 &   0.696(2) & 0.269(3)&&& \\
&&&     0.1464   & 5250 &   0.548(4) & 0.248(2)&&& \\
\hline
$16^3{\times}32$ & 1.95 & 1.53 &
        0.1375   & 7000 &   0.805(1) & 0.204(1) & 2.187 & 1.439 & 0.2079(15)\\
&&&     0.1390   & 7000 &   0.751(1) & 0.193(2) & 2.214 & 1.431 & 0.1977(13)\\
&&&     0.1400   & 7000 &   0.688(1) & 0.181(1) & 2.247 & 1.422 & 0.1853(9) \\
&&&     0.1410   & 7000 &   0.586(3) & 0.170(1) & 2.281 & 1.412 & 0.1727(10)\\
&&&              &      &            &          & 2.334 & 1.398 & 0.1577(9) \\
\hline
$24^3{\times}48$ & 2.10 & 1.47 & 
        0.1357   & 4000 &   0.806(1) & 0.1342(6)&2.416 & 1.379 & 0.1359(7)\\ 
&&&     0.1367   & 4000 &   0.755(2) & 0.1259(5)&2.456 & 1.370 & 0.1266(13\\
&&&     0.1374   & 4000 &   0.691(3) & 0.1201(5)&2.487 & 1.363 & 0.1206(9)\\
&&&     0.1382   & 4000 &   0.576(3) & 0.1128(3)&2.528 & 1.355 & 0.1130(9)\\
&&&              &      &            &          &2.575 & 1.345 & 0.1065(7)\\
\hline
$24^3{\times}48$ & 2.20 & 1.44 &
        0.1351   & 2000 &   0.800(2) & 0.1049(2) &&&\\ 
&&&     0.1358   & 2000 &   0.752(3) & 0.1012(3) &&&\\
&&&     0.1363   & 2000 &   0.702(3) & 0.0977(3) &&&\\
&&&     0.1368   & 2000 &   0.637(6) & 0.0947(2) &&&\\
\end{tabular}
\caption{Overview of full QCD and quenched configurations. 
         The scale $a_{\sigma}$, fixed by assuming 
        $\protect \sqrt{\sigma}=440$~MeV, is shown to illustrate the matching 
         of scales 
        (measurement of $\sigma$ at $\beta=2.10$ is made for the first 
        2000 trajectories).  
         The quenched runs have 200 configurations for each $\beta$.}
\label{tab:param-all}
\end{center}
\end{table}
\begin{table}
\begin{center}
\begin{tabular}{c|ccccc}
  $\beta$ & \multicolumn{5}{c}{$1.95$} \\
  \hline
  $K_{sea}$ & 0.1375 & 0.1390 & 0.1400 & 0.1410 & physical\\
  \hline
  $\aMSbar(1/a)$ & 0.2241 & 0.2241 & 0.2241 & 0.2241 & 0.2241 \\
  \hline
  $K_c$ & 
  0.144221(10) & 0.143648(11) & 0.143214(13) & 0.142737(11) & 0.142065(13) \\
  \hline
  $a^{-1}_\rho$ [GeV] & 
  0.990(4) & 1.048(5) & 1.097(6) & 1.191(9) & 1.269(14)\\
  $K_l$ & 
  0.144056(10) & 0.143498(11) & 0.143073(13) & 0.142613(11) & 0.141998(12) \\
  $K_s(K)$ & 
  0.13998(4) & 0.13979(4) & 0.13960(4)& 0.13956(5)& 0.13928(6) \\
  $K_s(\Phi)$ & 
  0.13908(6) & 0.13896(7) &0.13885(7)& 0.13899(8) & 0.13863(8) \\
  \hline
  $a^{-1}_\Upsilon$ [GeV] &
  1.115(8) & 1.185(25) & 1.242(12) & 1.337(15) & 1.469(53) \\  
  $K_l$ & 0.144091(10) & 0.143530(11) & 0.143104(12) & 0.142638(10) & \\
  $K_s(K)$ & 0.140870(8) & 0.140608(9) & 0.140373(8) & 0.140190(10)&  \\
  $K_s(\Phi)$ & 0.13965(4) & 0.13949(4)& 0.13935(4) & 0.13940(5) & \\
  \hline
  \hline
  $\beta$ & \multicolumn{5}{c}{2.1} \\
  \hline
  $K_{sea}$ & 0.1357 & 0.1367 & 0.1374 & 0.1382 & physical \\
  \hline
  $\aMSbar(1/a)$ & 0.204 & 0.204 & 0.204 & 0.204 & 0.204 \\
  \hline
  $K_c$ & 
  0.139748(19) & 0.13954(3) & 0.139388(13) & 0.139238(10) & 0.139022(19) \\
  \hline
  $a^{-1}_\rho$ [GeV] & 
  1.435(15) & 1.529(12) & 1.579(16) & 1.670(24) & 1.789(35) \\   
  $K_l$ & 
  0.139652(19)& 0.13945(3) & 0.139306(11) & 0.139160(10) & 0.138967(18) \\
  $K_s(K)$ & 
  0.13730(5) & 0.13732(4) & 0.13727(4) & 0.13727(6) & 0.13726(7) \\
  $K_s(\Phi)$ & 
  0.13686(10) & 0.13701(8) & 0.13700(7) & 0.13701(10) & 0.13701(10) \\
  \hline
  $a^{-1}_\Upsilon$ [GeV] & 1.772(25) & & & 2.010(35) &  \\ 
  $K_l$ & 0.139684(18)& & & 0.139183(9) & \\
  $K_s(K)$ & 0.138111(10) & & & 0.137835(9) &  \\
  $K_s(\Phi)$ & 0.13740(6)& & & 0.13737(5) & \\
\end{tabular}
\end{center}
\caption{
  Lattice spacings and physical values of the valence
  hopping parameter $K$ for full QCD.}
\label{tab:lat_physical_dyn}
\end{table}
\begin{table}
\begin{center}
\begin{tabular}{c|cccc}
  $\beta$ & 2.187 & 2.281 & 2.334 & 2.575 \\
  \hline
  $\aMSbar(1/a)$ &  0.2242 &  0.2122 & 0.2062 & 0.1829 \\
  \hline
  $K_c$ & 0.141666(12) & 0.139587(15) & 0.138728(13) & 0.136116(8) \\
  \hline
  $a^{-1}_\rho$ [GeV] & 1.017(10) & 1.116(12) & 1.207(12) & 1.743(17)\\  
  $K_l$      & 0.141504(12) & 0.139451(14) & 0.138604(13) & 0.136036(7) \\
  $K_s(K)$   & 0.13747(8)   & 0.13609(8)   & 0.13552(6)   & 0.13409(4)  \\
  $K_s(\Phi)$& 0.13651(15)  & 0.13522(15)  & 0.13477(11)  & 0.13368(7)  \\
  \hline
  $a^{-1}_\Upsilon$ [GeV] & 1.197(19) & 1.397(16) &  &  \\ 
  $K_l$      & 0.141557 & 0.139510 & &  \\
  $K_s(K)$   & 0.138824  & 0.137351 &  &  \\
  $K_s(\Phi)$& 0.137421  & 0.136161  &  &  \\
\end{tabular}
\end{center}
\caption{
  Lattice spacings and physical values of the hopping
  parameter $K$ for the quenched QCD runs.
  }
\label{tab:lat_physical_que}
\end{table}
\begin{table}
  \begin{center}
    \begin{tabular}{cccccc}
      $\beta$ & $f_B$ (MeV) & 
      \multicolumn{2}{c}{$f_{B_s}$ (MeV)} & 
      \multicolumn{2}{c}{$f_{B_s}/f_B$}\\
      \multicolumn{2}{c}{} &
      $K$ input & $\Phi$ input & $K$ input & $\Phi$ input \\
      \hline
      \multicolumn{6}{c}{scale from $m_\rho$} \\
      \hline
      2.187&224(6) & 256(5) & 265(5) & 1.146(10) & 1.184(12) \\ 
      2.281&199(7) & 230(6)  & 239(6)  & 1.154(20) & 1.197(25) \\ 
      2.334& 185(6)& 217(5)  & 226(5) & 1.177(18)  & 1.222(23)  \\ 
      2.575&191(4) & 220(4)  & 226(5)  & 1.150(9)  & 1.184(12)  \\ 
      \hline
      \multicolumn{6}{c}{scale from $\Upsilon$} \\
      \hline
      2.187  & 269(8) & 295(8) & 309(8) & 1.094(6)  & 1.149(9)  \\
      2.281  & 257(9) & 282(8) & 297(7) & 1.096(11) & 1.154(18) \\
    \end{tabular}
  \end{center}
  \caption{
    Quenched  decay constants.
    }
  \label{tab:resquenched}
\end{table}
\begin{table}
  \begin{center}
    \begin{tabular}{cccccc}
      $K_{sea}$ & $f_B$ (MeV) &
      \multicolumn{2}{c}{$f_{B_s}$ (MeV)} &
      \multicolumn{2}{c}{$f_{B_s}/f_B$}\\
      \multicolumn{2}{c}{} &
      $K$ input & $\Phi$ input & $K$ input & $\Phi$ input \\
      \hline
      \multicolumn{6}{c}{scale from $m_\rho$} \\
      \hline
      \multicolumn{6}{c}{$\beta$ = 1.95}\\
      \hline
      0.1375 & 230(4) & 266(4) & 275(4) & 1.156(9)  & 1.194(11)  \\
      0.1390 & 233(7)  & 266(6)  & 275(6) & 1.144(14) & 1.179(17)  \\
      0.1400 & 220(5)  & 259(4) & 268(4)  & 1.174(10) & 1.215(13) \\
      0.1410 & 228(6)  & 266(5)  & 273(6)  & 1.167(15) &  1.200(18)  \\
      \hline
      \multicolumn{6}{c|}{$\beta$ = 2.1}\\
      \hline
      0.1357 & 193(5)& 225(5) & 232(5) & 1.166(12)  & 1.199(14)  \\
      0.1367 & 206(6)& 238(5) & 243(6) & 1.159(11)  & 1.184(13)  \\
      0.1374 & 197(5)& 231(6) & 235(6) & 1.172(11)  & 1.196(12)  \\
      0.1382 & 201(8)& 236(9) & 242(8) & 1.177(15)  & 1.203(17)  \\
      \hline
      \hline
      \multicolumn{6}{c}{scale from $\Upsilon$} \\
      \hline
      \multicolumn{6}{c}{$\beta$ = 1.95}\\
      \hline
      0.1375 & 264(5) & 295(4) & 309(4) & 1.118(7)  & 1.168(10)  \\
      0.1390 & 266(10) & 294(10) & 306(10) & 1.108(9)  & 1.154(13)  \\
      0.1400 & 254(7) & 286(6) & 300(6) & 1.130(7)  & 1.183(10)  \\
      0.1410 & 259(8) & 293(7) & 305(7) & 1.130(12) & 1.176(15)  \\
      \hline
      \multicolumn{6}{c}{$\beta$ = 2.1}\\
      \hline
      0.1357 & 250(7) & 276(7) & 289(7) & 1.104(9) & 1.155(13)  \\
      0.1382 & 249(11) & 279(11) & 290(11) & 1.118(11) & 1.162(14)  \\
    \end{tabular}
  \end{center}
  \caption{
    Partially quenched decay constants.
    }
  \label{tab:respartquenched}
\end{table}
\begin{table}
  \begin{center}
    \begin{tabular}{cccccc}
      $\beta$ & $f_B$ (MeV) &
      \multicolumn{2}{c}{$f_{B_s}$ (MeV)} &
      \multicolumn{2}{c}{$f_{B_s}/f_B$} \\
      \hline
      & & $K$ input & $\Phi$ input & $K$ input & $\Phi$ input \\
      \hline
      \multicolumn{6}{c}{scale from $m_\rho$} \\
      \hline
      1.95 & 222(6) & 261(5) & 268(6) & 1.176(14) & 1.212(17) \\  
      2.1 & 204(8) & 242(9) & 245(9) & 1.179(18) & 1.198(20) \\
      \hline
      \multicolumn{6}{c}{scale from $\Upsilon$} \\
      \hline
      1.95 & 254(8)  & 287(7)  & 300(7)  & 1.134(10) & 1.183(14) \\
      2.1  & 249(16) & 280(16) & 290(16) & 1.124(16) & 1.165(21) \\
    \end{tabular}
  \end{center}
  \caption{
    Decay constants in full QCD.
    }
  \label{tab:resdyn}
\end{table}
\begin{table}
  \begin{center}
    \begin{tabular}{c|cccc|ccc}
      &
      \multicolumn{4}{c|}{$N_f$ = 0 lattices}&
      \multicolumn{2}{c}{$N_f$ = 2 lattices}\\
      \hline
      $\beta$ & 2.187 & 2.281 & 2.334 & 2.575 & 1.95 & 2.1 \\
      \hline
      $O(a^2 \Lambda_{QCD}^2)$     &
      35 \% & 29 \% & 25 \% & 12 \% & 22 \% & 11 \%\\
      $O(\alpha_s a\Lambda_{QCD})$ &
      13 \% & 11 \% & 10 \% & 6 \% & 11 \% & 7 \%\\
      $O(\alpha_s^2)$              & 
      5 \% & 5 \% & 4 \% & 3 \% & 5 \% & 4 \%\\
      $O(\alpha_s \Lambda_{QCD}/M_b)$ & 
      3 \% & 3 \% & 3 \% & 2 \% & 3 \% & 3 \%\\
      $O(\alpha_s^2/(aM_b))$       & 
      1 \% & 1 \% & 1 \% & 1 \% & 1 \% & 2 \%\\
      $O(\Lambda_{QCD}^2/M_b^2)$   & 
      2 \% & 2 \% & 2 \% & 1 \% & 2 \% & 2 \%\\
      $O(\Lambda_{QCD}^2 a/M_b)$   & 
      8 \% & 7 \% & 7 \% & 4 \% & 6 \% & 4 \%\\
      \hline
      total error for $f_B$ and $f_{B_s}$ & 
      39 \% & 32 \% & 28 \% & 14 \% & 26 \% & 14 \% \\
      \hline
      total error for $f_{B_s}/f_B-1$     & 
      37 \% & 31 \% & 27 \% & 13 \% & 25 \% & 13 \% \\
 \end{tabular}
 \end{center}
\caption{
  Estimated systematic errors.  
  $\aMSbar (1/a)$ and $\Lambda_{QCD}$ = 600 MeV are used.
  }
\label{tab:syst_errors_600}
\end{table}
%


\newpage

\section*{Appendix}

\begin{table}
 \begin{center}
 \begin{tabular}{cccccc}
 \multicolumn{6}{c}{$\beta$ = 1.95}\\
 \hline
 $aM_0$ &
 $K_{val}=0.1375$ & $K_{val}=0.1390$ & $K_{val}=0.1400$ &
 $K_{val}=0.1410$ & $K_{val}=0.1415$ \\
 \hline
 \multicolumn{6}{c}{$K_{sea}$ = 0.1375}\\
 \hline
 2.4& 3.45(4) & 3.38(4) &  3.34(5) & 3.31(6) & 3.29(6)  \\ 
 2.9& 3.95(5) & 3.88(5) &  3.83(6) & 3.80(7) & 3.79(8)  \\
 3.4& 4.43(6) & 4.36(7) &  4.31(8) & 4.28(9) & 4.28(10) \\
 4.0& 5.01(7) & 4.93(9) &  4.88(10)& 4.85(12)& 4.85(14) \\
 4.8& 5.76(10)& 5.67(12)&  5.62(13)& 5.60(16)& 5.61(19) \\
 \hline
 \multicolumn{6}{c}{$K_{sea}$ = 0.1390}\\
 \hline
 2.4 & 3.41(6)  & 3.36(7)  & 3.34(9)  & 3.36(13) & 3.42(16)\\
 2.9 & 3.91(8)  & 3.86(10) & 3.85(12) & 3.90(17) & 3.99(23)\\
 3.4 & 4.41(10) & 4.36(13) & 4.35(16) & 4.43(24) & 4.56(32)\\
 4.0 & 4.99(13) & 4.94(17) & 4.94(22) & 5.07(33) & 5.27(45)\\
 4.8 & 5.75(19) & 5.69(25) & 5.72(32) & 5.92(49) & 6.25(69)\\
 \hline
 \multicolumn{6}{c}{$K_{sea}$ = 0.1400}\\
 \hline
 2.4 & 3.39(5) & 3.31(8)  & 3.28(11) & 3.28(11) & 3.28(14)\\
 2.9 & 3.92(7) & 3.83(11) & 3.82(15) & 3.82(15) & 3.82(20)\\
 3.4 & 4.44(9) & 4.36(15) & 4.35(21) & 4.35(21) & 4.38(27)\\
 4.0 & 5.06(13)& 4.99(21) & 5.01(30) & 5.01(30) & 5.07(39)\\
 4.8 & 6.01(21)& 5.81(32) & 5.89(46) & 5.89(46) & 6.03(62)\\
 \hline
 \multicolumn{6}{c}{$K_{sea}$ = 0.1410}\\
 \hline
 2.4 & 3.22(6)  & 3.13(7)  & 3.06(9)  & 2.98(11) & 2.99(19)\\
 2.6 & 3.68(8)  & 3.58(9)  & 3.50(11) & 3.51(19) & 3.48(24)\\
 2.9 & 4.12(9)  & 4.00(11) & 3.91(14) & 3.99(24) & 3.98(30)\\
 3.2 & 4.63(12) & 4.49(14) & 4.39(17) & 4.56(31) & 4.59(40)\\
 3.5 & 5.27(15) & 5.11(19) & 4.99(23) & 5.33(45) & 5.42(59)\\
 \hline
 \hline
 \multicolumn{6}{c}{$\beta$ = 2.1}\\
 \hline
 $aM_0$ & 
 $K_{val}=0.1357$ & $K_{val}=0.1367$ & $K_{val}=0.1374$ &
 $K_{val}=0.1382$ & $K_{val}=0.1385$ \\
 \hline
 \multicolumn{6}{c}{$K_{sea}$ = 0.1357}\\
 \hline
 2.4 & 3.37(9)  & 3.26(14) & 3.29(18) & 3.39(27) & 3.34(34)\\
 2.6 & 3.44(13) & 3.48(17) & 3.51(21) & 3.63(33) & 3.58(41)\\
 2.9 & 3.74(17) & 3.80(21) & 3.89(27) & 4.00(43) & 3.93(54)\\
 3.2 & 4.04(21) & 4.13(26) & 4.19(34) & 4.38(56) & 4.28(71)\\
 3.5 & 4.34(25) & 4.45(33) & 4.53(42) & 4.78(72) & 4.66(48)\\
 \hline
 \multicolumn{6}{c}{$K_{sea}$ = 0.1367}\\
 \hline
 2.4  & 3.04(9)  & 3.03(10) & 3.04(11) & 3.03(14) & 3.04(16)\\
 2.6  & 3.23(10) & 3.23(11) & 3.24(13) & 3.22(16) & 3.24(18)\\
 2.9  & 3.51(12) & 3.52(14) & 3.53(16) & 3.52(20) & 3.54(23)\\
 3.2  & 3.79(14) & 3.81(17) & 3.83(19) & 3.81(25) & 3.84(27)\\
 3.5  & 4.07(17) & 4.09(10) & 4.13(23) & 4.10(30) & 4.13(33)\\
 \hline
 \multicolumn{6}{c}{$K_{sea}$ = 0.1374}\\
 \hline
 2.4 & 3.13(9)  & 3.09(11) & 3.05(12) & 3.00(15) & 2.96(18)\\
 2.6 & 3.33(11) & 3.28(12) & 3.24(13) & 3.19(17) & 3.14(20)\\
 2.9 & 3.62(12) & 3.56(14) & 3.52(15) & 3.46(21) & 3.42(24)\\
 3.2 & 3.91(14) & 3.84(16) & 3.79(18) & 3.74(24) & 3.70(29)\\
 3.5 & 4.19(16) & 4.11(18) & 4.05(20) & 4.01(28) & 3.98(34)\\
 \hline
 \multicolumn{6}{|c|}{$K_{sea}$ = 0.1382}\\
 \hline
 2.4 & 2.95(8)  & 2.93(10) & 2.93(13) & 2.92(18) & 2.94(22)\\
 2.6 & 3.12(8)  & 3.11(11) & 3.11(14) & 3.12(20) & 3.13(25)\\
 2.9 & 3.39(10) & 3.38(13) & 3.39(16) & 3.40(24) & 3.42(29)\\
 3.2 & 3.65(11) & 3.65(15) & 3.67(19) & 3.68(28) & 3.69(34)\\
 3.5 & 3.91(11) & 3.92(17) & 3.94(22) & 3.95(32) & 3.96(40)\\
 \hline
\end{tabular}
\end{center}
\caption{
  Kinetic masses in lattice units measured from the
  dispersion relation in full QCD.
  }
\label{tab:Mkin_dyn}
\end{table}
\begin{table}
\begin{center}
\begin{tabular}{cccccc}
 \multicolumn{6}{c}{$\beta = 2.187$}\\
 \hline
 $aM_0$ & 
 $K=0.1351$ & $K=0.1365$ & $K=0.1375$ & $K=0.1385$ & $K=0.1390$\\
 \hline
 2.4&  3.36(8)  & 3.29(8)  & 3.26(9)  & 3.23(11) & 3.24(12) \\
 2.9&  3.84(9)  & 3.79(10) & 3.76(12) & 3.73(14) & 3.71(15) \\
 3.4&  4.33(11) & 4.28(13) & 4.25(15) & 4.22(17) & 4.21(19) \\
 4.0&  4.91(14) & 4.86(17) & 4.83(19) & 4.82(23) & 4.82(25) \\
 4.8&  5.65(19) & 5.61(23) & 5.60(27) & 5.62(32) & 5.65(35) \\
 \hline
 \multicolumn{6}{c}{$\beta = 2.281$}\\
 \hline
 $aM_0$ & 
 $K=0.1343$ & $K=0.1357$ & $K=0.1367$ & $K=0.1377$ & $K=0.1383$\\
 \hline
 2.4&  3.35(7)  &  3.28(8)  & 3.22(9)  & 3.17(11) & 3.17(14)\\
 2.9&  3.88(9)  &  3.79(11) & 3.72(12) & 3.66(15) & 3.64(18)\\
 3.4&  4.40(12) &  4.29(14) & 4.20(16) & 4.13(19) & 4.10(23)\\
 4.0&  5.01(16) &  4.88(18) & 4.78(21) & 4.67(25) & 4.63(30)\\
 4.8&  5.82(23) &  5.66(26) & 5.51(29) & 5.38(35) & 5.32(41)\\
 \hline
 \multicolumn{6}{c}{$\beta = 2.334$}\\
 \hline
 $aM_0$ & 
 $K=0.1337$ & $K=0.1349$ & $K=0.1358$ & $K=0.1368$ & $K=0.1374$\\
 \hline
 2.4& 3.17(7)  & 3.11(7)  & 3.08(8)  & 3.04(9)  & 3.03(11)\\
 2.9& 3.64(9)  & 3.58(10) & 3.54(11) & 3.51(12) & 3.47(15)\\
 3.4& 4.10(11) & 4.04(12) & 4.00(14) & 3.96(16) & 3.90(19)\\
 4.0& 4.64(15) & 4.57(16) & 4.52(18) & 4.41(20) & 4.39(24)\\
 4.8& 5.33(20) & 5.25(22) & 5.19(24) & 5.03(27) & 5.00(32)\\
 \hline
 \multicolumn{6}{c}{$\beta = 2.575$}\\
 \hline
 $aM_0$ & 
 $K=0.1329$ & $K=0.1337$ & $K=0.1344$ & $K=0.1351$ & $K=0.1353$\\
 \hline
 2.4& 2.77(09) & 2.71(11) & 2.65(13) & 2.57(17) & 2.55(19) \\
 2.6& 2.94(10) & 2.87(12) & 2.89(15) & 2.73(19) & 2.69(22) \\
 2.9& 3.18(12) & 3.11(14) & 3.04(18) & 2.95(23) & 2.91(26) \\
 3.2& 3.43(14) & 3.35(17) & 3.26(21) & 3.17(28) & 3.13(31) \\
 3.5& 3.67(16) & 3.58(20) & 3.49(24) & 3.38(33) & 3.34(36) \\
\end{tabular}
\end{center}
\caption{
  Kinetic masses in lattice units measured from the
  dispersion relation in the quenched case.
  }
\label{tab:Mkin_quenched}
\end{table}
\begin{table}
\begin{center}
\begin{tabular}{cccccc}
 \multicolumn{6}{c}{$\beta = 1.95$}\\
 \hline
 $aM^0$ &
 $K_{val}=0.1375$ & $K_{val}=0.1390$ & $K_{val}=0.1400$ &
 $K_{val}=0.1410$ & $K_{val}=0.1415$ \\
 \hline
 \multicolumn{6}{c}{$K_{sea} = 0.1375$}\\
 \hline
 2.4  & 0.712(1) & 0.679(1) & 0.657(2) & 0.635(2) & 0.624(2)\\
 2.9  & 0.730(1) & 0.697(1) & 0.675(2) & 0.654(2) & 0.643(2)\\
 3.4  & 0.741(1) & 0.709(2) & 0.688(2) & 0.667(2) & 0.656(2)\\
 4.0  & 0.750(1) & 0.719(2) & 0.698(2) & 0.677(2) & 0.667(2)\\
 4.8  & 0.757(2) & 0.727(2) & 0.706(2) & 0.686(2) & 0.676(3)\\
 \hline
 \multicolumn{6}{c}{$K_{sea} = 0.1390$}\\
 \hline
 2.4  & 0.694(2) & 0.661(2) & 0.639(2) & 0.617(2) & 0.607(3)\\
 2.9  & 0.712(2) & 0.679(2) & 0.658(2) & 0.637(3) & 0.627(3)\\
 3.4  & 0.724(2) & 0.692(2) & 0.671(2) & 0.650(3) & 0.640(3)\\
 4.0  & 0.734(2) & 0.703(2) & 0.682(2) & 0.662(3) & 0.652(3)\\
 4.8  & 0.743(2) & 0.712(2) & 0.692(3) & 0.672(3) & 0.662(4)\\
 \hline
 \multicolumn{6}{c}{$K_{sea} = 0.1400$}\\
 \hline
 2.4  & 0.677(1) & 0.641(2) & 0.618(2) & 0.594(2) & 0.582(3)\\
 2.9  & 0.694(2) & 0.659(2) & 0.636(2) & 0.613(3) & 0.601(3)\\
 3.4  & 0.705(2) & 0.671(2) & 0.649(2) & 0.626(3) & 0.614(3)\\
 4.0  & 0.715(2) & 0.681(2) & 0.659(2) & 0.636(3) & 0.624(3)\\
 4.8  & 0.723(2) & 0.690(2) & 0.667(3) & 0.645(3) & 0.634(4)\\
 \hline
 \multicolumn{6}{c}{$K_{sea} = 0.1410$}\\
 \hline
 2.4& 0.656(2) & 0.620(2) & 0.596(3) & 0.573(3) & 0.561(3)\\
 2.9& 0.673(2) & 0.637(2) & 0.614(3) & 0.591(3) & 0.580(4)\\
 3.4& 0.684(2) & 0.649(2) & 0.627(3) & 0.604(3) & 0.592(4)\\
 4.0& 0.693(2) & 0.659(3) & 0.637(3) & 0.614(4) & 0.603(4)\\
 4.8& 0.701(2) & 0.668(3) & 0.646(3) & 0.624(4) & 0.612(4)\\
 \hline
 \hline
 \multicolumn{6}{c}{$\beta = 2.1$}\\
 \hline
 $aM^0$ &
 $K_{val}=0.1357$ & $K_{val}=0.1367$ & $K_{val}=0.1374$ &
 $K_{val}=0.1382$ & $K_{val}=0.1385$\\
 \hline
 \multicolumn{6}{c}{$K_{sea} = 0.1357$}\\
 \hline
 2.4   &0.585(2) &0.559(2) &0.540(2) &0.519(3) &0.510(3)\\
 2.6   &0.592(2) &0.566(2) &0.547(2) &0.526(3) &0.517(3)\\
 2.9   &0.600(2) &0.574(2) &0.555(2) &0.534(3) &0.526(3)\\
 3.2   &0.606(2) &0.580(2) &0.562(2) &0.541(3) &0.532(3)\\
 3.5   &0.611(2) &0.585(2) &0.567(2) &0.546(3) &0.538(3)\\
 \hline
 \multicolumn{6}{c}{$K_{sea} = 0.1367$}\\
 \hline
 2.4   & 0.575(2) & 0.548(2) & 0.529(2) & 0.507(3) & 0.499(3)\\
 2.6   & 0.582(2) & 0.555(2) & 0.536(2) & 0.514(3) & 0.506(3)\\
 2.9   & 0.590(2) & 0.563(2) & 0.544(2) & 0.523(3) & 0.515(3)\\
 3.2   & 0.596(2) & 0.570(2) & 0.551(2) & 0.530(3) & 0.522(3)\\
 3.5   & 0.601(2) & 0.575(2) & 0.556(3) & 0.536(3) & 0.528(4)\\
 \hline
 \multicolumn{6}{c}{$K_{sea} = 0.1374$}\\
 \hline
 2.4  & 0.564(2) & 0.536(2) & 0.516(2) & 0.494(3) & 0.485(3)\\
 2.6  & 0.570(2) & 0.542(2) & 0.523(2) & 0.500(3) & 0.492(3)\\
 2.9  & 0.578(2) & 0.550(2) & 0.531(2) & 0.509(3) & 0.500(3)\\
 3.2  & 0.584(2) & 0.557(2) & 0.538(2) & 0.515(3) & 0.507(3)\\
 3.5  & 0.589(2) & 0.562(2) & 0.543(2) & 0.521(3) & 0.512(3)\\
 \hline
 \multicolumn{6}{c}{$K_{sea} = 0.1382$}\\
 \hline
 2.4  & 0.553(2) & 0.524(2) & 0.505(3) & 0.483(3) & 0.476(4)\\
 2.6  & 0.559(2) & 0.531(2) & 0.511(3) & 0.490(3) & 0.483(4)\\
 2.9  & 0.567(2) & 0.539(3) & 0.519(3) & 0.498(3) & 0.491(4)\\
 3.2  & 0.574(2) & 0.546(3) & 0.526(3) & 0.505(3) & 0.498(4)\\
 3.5  & 0.579(2) & 0.551(3) & 0.532(3) & 0.510(4) & 0.503(4)\\
\end{tabular}
\end{center}
\caption{
  Simulation energies in full QCD, in lattice units.
  }
\label{tab:Esim_dyn}
\end{table}

\begin{table}
\begin{center}
\begin{tabular}{cccccc}
 \multicolumn{6}{c}{$\beta = 2.187$}\\
 \hline
 $aM_0$ &
 $K=0.1351$ & $K=0.1365$ & $K=0.1375$ & $K=0.1385$ & $K=0.1390$\\
 \hline
 2.4 &0.707(2) &0.675(2) &0.652(3) &0.629(3) &0.617(3)\\
 2.6 &0.724(2) &0.693(2) &0.671(3) &0.649(3) &0.637(4)\\
 2.9 &0.736(2) &0.706(2) &0.684(3) &0.662(3) &0.651(4)\\
 3.2 &0.746(2) &0.716(3) &0.694(3) &0.673(3) &0.662(4)\\
 3.5 &0.754(2) &0.725(3) &0.704(3) &0.683(3) &0.672(4)\\
 \hline
 \multicolumn{6}{c}{$\beta = 2.281$}\\
 \hline
 $aM_0$ &
 $K=0.1343$ & $K=0.1357$ & $K=0.1367$ & $K=0.1377$ & $K=0.1383$\\
 \hline
 2.4 & 0.648(2) & 0.614(3) & 0.590(3) & 0.566(4) & 0.551(4)\\
 2.9 & 0.664(3) & 0.631(3) & 0.608(3) & 0.584(4) & 0.569(5)\\
 3.4 & 0.674(3) & 0.642(3) & 0.619(4) & 0.597(5) & 0.582(5)\\
 4.0 & 0.683(3) & 0.652(3) & 0.629(4) & 0.607(5) & 0.593(5)\\
 4.8 & 0.691(3) & 0.660(3) & 0.639(4) & 0.617(5) & 0.604(5)\\
 \hline
 \multicolumn{6}{c}{$\beta = 2.334$}\\
 \hline
 $aM_0$ &
 $K=0.1337$ & $K=0.1349$ & $K=0.1358$ & $K=0.1368$ & $K=0.1374$\\
 \hline
 2.4 &0.623(2) &0.592(3) &0.569(3) &0.544(4) &0.529(5)\\
 2.9 &0.638(2) &0.608(3) &0.586(3) &0.561(4) &0.546(5)\\
 3.4 &0.648(2) &0.619(3) &0.597(3) &0.572(4) &0.558(5)\\
 4.0 &0.657(3) &0.628(3) &0.606(3) &0.582(4) &0.568(5)\\
 4.8 &0.665(3) &0.636(3) &0.615(4) &0.591(5) &0.577(6)\\
 \hline
 \multicolumn{6}{c}{$\beta = 2.575$}\\
 \hline
 $aM_0$ &
 $K=0.1329$ & $K=0.1337$ & $K=0.1344$ & $K=0.1351$ & $K=0.1353$\\
 \hline
 2.4 &  0.501(2) &  0.478(2) &  0.459(2) &  0.440(3) &  0.435(3)\\
 2.9 &  0.506(2) &  0.484(2) &  0.464(3) &  0.446(3) &  0.441(3)\\
 3.4 &  0.513(2) &  0.491(2) &  0.472(3) &  0.454(3) &  0.449(4)\\
 4.0 &  0.518(2) &  0.496(2) &  0.478(3) &  0.460(3) &  0.455(4)\\
 4.8 &  0.523(2) &  0.501(2) &  0.482(3) &  0.465(3) &  0.460(4)\\
 \end{tabular}
 \end{center}
\caption{
  Quenched simulation energies, in lattice units.
  }
\label{tab:Esim_quenched}
\end{table}
\begin{table}
\begin{center}
\begin{tabular}{cccccc}
 \multicolumn{6}{c}{$\beta = 1.95$}\\
 \hline
 $aM_0$ &
 $K_{val}=0.1375$ & $K_{val}=0.1390$ & $K_{val}=0.1400$ &
 $K_{val}=0.1410$ & $K_{val}=0.1415$ \\
 \hline
 \multicolumn{6}{c}{$K_{sea} = 0.1375$}\\
 \hline
 2.4 & 3.08(15) & 3.04(15) & 3.02(15) & 3.00(15) & 2.99(15)\\
 2.9 & 3.64(18) & 3.60(18) & 3.58(18) & 3.56(18) & 3.55(18)\\
 3.4 & 4.18(21) & 4.15(21) & 4.13(21) & 4.11(21) & 4.10(21)\\
 4.0 & 4.83(24) & 4.80(24) & 4.78(24) & 4.76(24) & 4.75(24)\\
 4.8 & 5.69(28) & 5.66(28) & 5.64(28) & 5.62(28) & 5.61(28)\\
 \hline
 \multicolumn{6}{c}{$K_{sea} = 0.1390$}\\
 \hline
 2.4 & 3.06(15) & 3.02(15) & 3.00(15) & 2.98(15) & 2.97(15)\\
 2.9 & 3.62(18) & 3.59(18) & 3.56(18) & 3.54(18) & 3.53(18)\\
 3.4 & 4.17(21) & 4.14(21) & 4.11(21) & 4.09(20) & 4.08(20)\\
 4.0 & 4.82(24) & 4.79(24) & 4.76(24) & 4.74(24) & 4.73(24)\\
 4.8 & 5.67(28) & 5.64(28) & 5.62(28) & 5.60(28) & 5.59(28)\\
 \hline
 \multicolumn{6}{c}{$K_{sea} = 0.1400$}\\
 \hline
 2.4 & 3.04(15) & 3.00(15) & 2.98(15) & 2.96(15) & 2.95(15)\\
 2.9 & 3.60(18) & 3.57(18) & 3.54(18) & 3.52(18) & 3.51(18)\\
 3.4 & 4.15(21) & 4.12(21) & 4.09(20) & 4.07(20) & 4.06(20)\\
 4.0 & 4.80(24) & 4.76(24) & 4.74(24) & 4.72(24) & 4.71(24)\\
 4.8 & 5.65(28) & 5.62(28) & 5.60(28) & 5.57(28) & 5.56(28)\\
 \hline
 \multicolumn{6}{c}{$K_{sea} = 0.1410$}\\
 \hline
 2.4  & 3.02(15) & 2.98(15) & 2.96(15) & 2.94(15) & 2.92(15)\\
 2.6  & 3.58(18) & 3.54(18) & 3.52(18) & 3.50(18) & 3.49(17)\\
 2.9  & 4.13(21) & 4.09(20) & 4.07(20) & 4.05(20) & 4.04(20)\\
 3.2  & 4.78(24) & 4.74(24) & 4.72(24) & 4.70(24) & 4.69(23)\\
 3.5  & 5.63(28) & 5.60(28) & 5.58(28) & 5.55(28) & 5.54(28)\\
 \hline
 \hline
 \multicolumn{6}{c}{$\beta = 2.1$}\\
 \hline
 $aM_0$ &
 $K_{val}=0.1357$ & $K_{val}=0.1367$ & $K_{val}=0.1374$ &
 $K_{val}=0.1382$ & $K_{val}=0.1385$ \\
 \hline
 \multicolumn{6}{c}{$K_{sea} = 0.1357$}\\
 \hline
 2.4 & 2.95(12) & 2.93(12) & 2.91(12) & 2.89(12) & 2.88(12)\\
 2.6 & 3.18(13) & 3.15(13) & 3.13(13) & 3.11(12) & 3.10(12)\\
 2.9 & 3.51(14) & 3.48(14) & 3.46(14) & 3.44(14) & 3.43(14)\\
 3.2 & 3.83(15) & 3.81(15) & 3.79(15) & 3.77(15) & 3.76(15)\\
 3.5 & 4.16(17) & 4.13(17) & 4.11(16) & 4.09(16) & 4.08(16)\\
 \hline
 \multicolumn{6}{c}{$K_{sea} = 0.1367$}\\
 \hline
 2.4 & 2.94(12) & 2.91(12) & 2.90(12) & 2.87(12) & 2.87(12)\\
 2.6 & 3.17(13) & 3.14(13) & 3.12(12) & 3.10(12) & 3.09(12)\\  
 2.9 & 3.50(14) & 3.47(14) & 3.45(14) & 3.43(14) & 3.42(14)\\
 3.2 & 3.82(15) & 3.80(15) & 3.78(15) & 3.76(15) & 3.75(15)\\
 3.5 & 4.15(17) & 4.12(16) & 4.10(16) & 4.08(16) & 4.07(16)\\
 \hline
 \multicolumn{6}{c}{$K_{sea} = 0.1374$}\\
 \hline
 2.4 & 2.93(12) & 2.90(12) & 2.88(12) & 2.86(11) & 2.85(11)\\
 2.6 & 3.15(13) & 3.13(13) & 3.11(12) & 3.08(12) & 3.08(12)\\
 2.9 & 3.48(14) & 3.46(14) & 3.44(14) & 3.41(14) & 3.41(14)\\
 3.2 & 3.81(15) & 3.78(15) & 3.76(15) & 3.74(15) & 3.73(15)\\
 3.5 & 4.14(17) & 4.11(16) & 4.09(16) & 4.07(16) & 4.06(16)\\
 \hline
 \multicolumn{6}{c}{$K_{sea} = 0.1382$}\\
 \hline
 2.4 & 2.92(12) & 2.89(12) & 2.87(11) & 2.85(11) & 2.84(11)\\
 2.6 & 3.14(13) & 3.11(12) & 3.09(12) & 3.07(12) & 3.07(12)\\
 2.9 & 3.47(14) & 3.44(14) & 3.43(14) & 3.40(14) & 3.40(14)\\
 3.2 & 3.80(15) & 3.77(15) & 3.75(15) & 3.73(15) & 3.72(15)\\
 3.5 & 4.13(17) & 4.10(16) & 4.08(16) & 4.06(16) & 4.05(16)\\
 \hline
 \end{tabular}
 \end{center}
\caption{
  Meson masses with the perturbative mass shifts in full
  QCD, in lattice units.
  } 
\label{tab:Mpert_dyn}
\end{table}

\begin{table}
\begin{center}
\begin{tabular}{cccccc}
 \multicolumn{6}{c}{$\beta = 2.187$}\\
 \hline
 $aM_0$ &
 $K=0.1351$ & $K=0.1365$ & $K=0.1375$ & $K=0.1385$ & $K=0.1390$\\
 \hline
 2.4 & 3.07(15) & 3.04(15) & 3.02(15) & 2.99(15) & 2.98(15)\\ 
 2.9 & 3.63(18) & 3.60(18) & 3.58(18) & 3.55(18) & 3.54(18)\\
 3.4 & 4.18(21) & 4.15(21) & 4.13(21) & 4.11(21) & 4.09(20)\\
 4.0 & 4.83(24) & 4.80(24) & 4.78(24) & 4.76(24) & 4.75(24)\\
 4.8 & 5.68(28) & 5.65(28) & 5.63(28) & 5.61(28) & 5.60(28)\\
 \hline
 \multicolumn{6}{c}{$\beta = 2.281$}\\
 \hline
 $aM_0$ &
 $K=0.1343$ & $K=0.1357$ & $K=0.1367$ & $K=0.1377$ & $K=0.1383$\\
 \hline
 2.4 & 3.01(15) & 2.98(15) & 2.96(15) & 2.93(15) & 2.92(15)\\
 2.9 & 3.57(18) & 3.54(18) & 3.51(18) & 3.49(17) & 3.48(17)\\
 3.4 & 4.12(21) & 4.08(20) & 4.06(20) & 4.04(20) & 4.02(20)\\
 4.0 & 4.76(24) & 4.73(24) & 4.71(24) & 4.69(23) & 4.67(23)\\
 4.8 & 5.61(28) & 5.58(28) & 5.56(28) & 5.54(28) & 5.53(28)\\
 \hline
 \multicolumn{6}{c}{$\beta = 2.334$}\\
 \hline
 $aM_0$ &
 $K=0.1337$ & $K=0.1349$ & $K=0.1358$ & $K=0.1368$ & $K=0.1374$\\
 \hline
 2.4 &2.99(12) &2.96(12) &2.94(12) &2.91(12) &2.90(12)\\
 2.9 &3.54(14) &3.51(14) &3.49(14) &3.47(14) &3.45(14)\\
 3.4 &4.09(16) &4.06(16) &4.04(16) &4.01(16) &4.00(16)\\
 4.0 &4.73(19) &4.70(19) &4.68(19) &4.66(19) &4.64(19)\\
 4.8 &5.58(22) &5.56(22) &5.53(22) &5.51(22) &5.50(22)\\
 \hline
 \multicolumn{6}{c}{$\beta = 2.575$}\\
 \hline
 $aM_0$ &
 $K=0.1329$ & $K=0.1337$ & $K=0.1344$ & $K=0.1351$ & $K=0.1353$\\
 \hline
 2.4 & 2.87(9) & 2.85(9) & 2.83(8) & 2.81(8) & 2.81(8)\\
 2.6 & 3.09(9) & 3.07(9) & 3.05(9) & 3.03(9) & 3.03(9)\\
 2.9 & 3.42(10) & 3.40(10) & 3.38(10) & 3.36(10) & 3.36(10)\\
 3.2 & 3.75(11) & 3.72(11) & 3.70(11) & 3.69(11) & 3.68(12)\\
 3.5 & 4.07(12) & 4.05(12) & 4.03(12) & 4.01(12) & 4.00(12)\\
 \end{tabular}
 \end{center}
\caption{
  Quenched meson masses with perturbative mass shifts, in
  lattice units.
  } 
\label{tab:Mpert_que}
\end{table}
\begin{table}
  \begin{center}
    \begin{tabular}{cccc}
      $\beta$ & $K_{sea}$ & 
      scale from $\rho$ & 
      scale from $\Upsilon$ \\ 
      \hline
      1.95 & 0.1375 & 4.53(12) & 4.43(13) \\
           & 0.1390 & 4.29(26) & 4.14(28) \\
           & 0.1400 & 4.26(21) & 4.18(23) \\
           & 0.1410 & 4.88(23) & 4.77(23) \\
      \hline
      2.1  & 0.1357 & 4.08(34) & 3.95(30) \\
           & 0.1367 & 4.45(24) &          \\
           & 0.1374 & 4.41(25) &          \\
           & 0.1382 & 4.65(42) & 4.44(43) \\
    \end{tabular}
  \end{center}
  \caption{
    Bare $b$ quark masses $M_{0b}$, in GeV, for the full QCD
    lattices. 
    Results with two different scale determinations,
    $m_{\rho}$ and $\Upsilon$ are listed for each sea quark
    mass. 
    }
  \label{tab:Mb0_PQ}
\end{table}
\begin{table}
  \begin{center}
    \begin{tabular}{ccc}
      $\beta$ & 
      scale from $\rho$ & 
      scale from $\Upsilon$ \\ 
      \hline
      2.187 & 4.55(24) & 4.42(25) \\
      2.281 & 4.45(23) & 4.32(24) \\
      2.334 & 4.54(27) &          \\
      2.575 & 5.15(39) &           \\
    \end{tabular}
  \end{center}
  \caption{
    Bare $b$ quark masses $M_{0b}$, in GeV, for the quenched
    lattices. 
    Results with two different scale determinations,
    $m_{\rho}$ and $\Upsilon$ are listed for each sea quark
    mass. 
    }
  \label{tab:Mb0_que}
\end{table}
\begin{table}
 \begin{center}
 \begin{tabular}{cccccc}
 \multicolumn{6}{c}{$a^{3/2}(f\sqrt{M})^{(0)}$}\\
 \hline
 $aM_0$&
 $K_{val}=0.1375$ & $K_{val}=0.1390$ & $K_{val}=0.1400$ &
 $K_{val}=0.1410$ & $K_{val}=0.1415$\\
 \hline
 \multicolumn{6}{c}{$K_{sea} = 0.1375$}\\
 \hline
 2.4&0.567(5)&0.576(5)&0.595(4)&0.613(4)&0.640(4)\\
 2.9&0.595(6)&0.606(5)&0.626(5)&0.646(4)&0.676(4)\\
 3.4&0.619(6)&0.630(6)&0.652(5)&0.674(5)&0.705(5)\\
 4.0&0.643(7)&0.655(6)&0.678(6)&0.702(5)&0.736(5)\\
 4.8&0.669(7)&0.682(7)&0.707(6)&0.732(6)&0.768(6)\\
\hline
 \multicolumn{6}{c}{$K_{sea} = 0.1390$}\\
 \hline
 2.4&0.534(7)&0.543(6)&0.561(5)&0.579(5)&0.606(5)\\
 2.9&0.561(7)&0.571(7)&0.591(6)&0.611(6)&0.640(5)\\
 3.4&0.586(10)&0.597(9)&0.619(8)&0.641(7)&0.673(7)\\
 4.0&0.612(11)&0.623(10)&0.646(9)&0.670(8)&0.704(7)\\
 4.8&0.641(13)&0.653(11)&0.678(10)&0.702(09)&0.739(8)\\
 \hline
 \multicolumn{6}{c}{$K_{sea} = 0.1400$}\\
 \hline
 2.4&0.484(6)&0.496(5)&0.518(4)&0.539(4)&0.570(4)\\
 2.9&0.508(6)&0.520(6)&0.544(5)&0.567(5)&0.600(4)\\
 3.4&0.527(7)&0.540(6)&0.566(5)&0.590(5)&0.626(5)\\
 4.0&0.546(8)&0.560(7)&0.588(6)&0.614(5)&0.652(5)\\
 4.8&0.567(9)&0.582(8)&0.612(7)&0.640(6)&0.680(6)\\
 \hline
 \multicolumn{6}{c}{$K_{sea} = 0.1410$}\\
 \hline
 2.4&0.429(7)&0.441(6)&0.462(5)&0.484(5)&0.514(5)\\
 2.9&0.449(9)&0.462(7)&0.486(6)&0.509(6)&0.542(5)\\
 3.4&0.465(9)&0.479(8)&0.505(7)&0.530(6)&0.565(6)\\
 4.0&0.482(11)&0.497(9)&0.525(8)&0.551(7)&0.587(6)\\
 4.8&0.501(12)&0.517(10)&0.546(8)&0.573(8)&0.612(7)\\
 \hline
 \hline
 \multicolumn{6}{c}{$a^{3/2}(f\sqrt{M})^{(1)}$}\\
 \hline
 $aM_0$ &
 $K_{val}=0.1375$ & $K_{val}=0.1390$ & $K_{val}=0.1400$ &
 $K_{val}=0.1410$ & $K_{val}=0.1415$\\
 \hline
 \multicolumn{6}{c}{$K_{sea} = 0.1375$}\\
 \hline
 2.4&$-$0.0817(12)&$-$0.0823(12)&$-$0.0836(10)&$-$0.0850(10)&$-$0.0869(9)\\
 2.9&$-$0.0730(12)&$-$0.0735(11)&$-$0.0748(10)&$-$0.0761(9)&$-$0.0779(8)\\
 3.4&$-$0.0661(11)&$-$0.0666(11)&$-$0.0677(9)&$-$0.0689(9)&$-$0.0706(8)\\
 4.0&$-$0.0594(11)&$-$0.0599(10)&$-$0.0609(9)&$-$0.0620(8)&$-$0.0636(7)\\
 4.8&$-$0.0524(10)&$-$0.0528(9)&$-$0.0538(8)&$-$0.0547(8)&$-$0.0562(7)\\
 \hline
 \multicolumn{6}{c}{$K_{sea} = 0.1390$}\\
 \hline
 2.4&$-$0.0754(11)&$-$0.0761(10)&$-$0.0775(9)&$-$0.0789(8)&$-$0.0810(8)\\
 2.9&$-$0.0672(10)&$-$0.0678(9) &$-$0.0691(8)&$-$0.0705(8)&$-$0.0724(7)\\
 3.4&$-$0.0614(11)&$-$0.0619(10)&$-$0.0631(9)&$-$0.0643(8)&$-$0.0662(7)\\
 4.0&$-$0.0552(11)&$-$0.0557(10)&$-$0.0568(9)&$-$0.0580(8)&$-$0.0597(7)\\
 4.8&$-$0.0488(11)&$-$0.0493(10)&$-$0.0503(9)&$-$0.0514(8)&$-$0.0530(7)\\
\hline
 \multicolumn{6}{c}{$K_{sea} = 0.1400$}\\
 \hline
 2.4&$-$0.0681(8)&$-$0.0692(8)&$-$0.0712(7)&$-$0.0730(6)&$-$0.0755(6)\\
 2.9&$-$0.0604(8)&$-$0.0614(7)&$-$0.0633(6)&$-$0.0650(6)&$-$0.0674(5)\\
 3.4&$-$0.0544(8)&$-$0.0553(7)&$-$0.0570(6)&$-$0.0587(5)&$-$0.0609(5)\\
 4.0&$-$0.0486(8)&$-$0.0494(7)&$-$0.0510(6)&$-$0.0525(5)&$-$0.0546(5)\\
 4.8&$-$0.0425(7)&$-$0.0433(6)&$-$0.0447(5)&$-$0.0461(5)&$-$0.0481(5)\\
 \hline
 \multicolumn{6}{c}{$K_{sea} = 0.1410$}\\
 \hline
 2.4&$-$0.0607(12)&$-$0.0616(10)&$-$0.0637(8)&$-$0.0656(8)&$-$0.0682(7)\\
 2.9&$-$0.0537(12)&$-$0.0545(10)&$-$0.0565(8)&$-$0.0583(7)&$-$0.0607(7)\\
 3.4&$-$0.0482(11)&$-$0.0490(9) &$-$0.0508(8)&$-$0.0525(7)&$-$0.0547(6)\\
 4.0&$-$0.0430(10)&$-$0.0437(9) &$-$0.0453(7)&$-$0.0469(7)&$-$0.0490(6)\\
 4.8&$-$0.0376(10)&$-$0.0382(8) &$-$0.0397(7)&$-$0.0411(6)&$-$0.0430(6)\\
 \end{tabular}
 \end{center}
\caption{Decay matrix elements at $\beta = 1.95$.}
\label{tab:frootm_beta1.95}
\end{table}
\begin{table}
 \begin{center}
 \begin{tabular}{cccccc}
 \multicolumn{6}{c}{$a^{3/2}(f\sqrt{M})^{(0)}$}\\
 \hline
 $aM_0$ &
 $K_{val}=0.1357$ & $K_{val}=0.1367$ & $K_{val}=0.1374$ &
 $K_{val}=0.1382$ & $K_{val}=0.1385$\\
 \hline
 \multicolumn{6}{c}{$K_{sea} = 0.1357$}\\
 \hline
 2.4&0.317(5)&0.324(4)&0.341(4)&0.355(4)&0.376(4)\\
 2.6&0.322(5)&0.329(4)&0.346(4)&0.361(4)&0.383(4)\\
 2.9&0.328(5)&0.336(4)&0.354(4)&0.369(4)&0.392(4)\\
 3.2&0.334(5)&0.342(4)&0.360(4)&0.376(4)&0.399(4)\\
 3.5&0.339(5)&0.347(4)&0.366(4)&0.382(4)&0.406(4)\\
 \hline
 \multicolumn{6}{c}{$K_{sea} = 0.1367$}\\
 \hline
 2.4&0.299(5)&0.306(5)&0.324(4)&0.339(4)&0.360(3)\\
 2.6&0.304(5)&0.311(5)&0.329(4)&0.345(4)&0.366(4)\\
 2.9&0.311(5)&0.318(5)&0.337(4)&0.353(4)&0.375(4)\\
 3.2&0.317(6)&0.324(5)&0.343(4)&0.359(4)&0.382(4)\\
 3.5&0.322(6)&0.329(5)&0.348(5)&0.365(4)&0.389(4)\\
 \hline
 \multicolumn{6}{c}{$K_{sea} = 0.1374$}\\
 \hline
 2.4&0.275(4)&0.282(4)&0.300(3)&0.316(3)&0.337(4)\\
 2.6&0.278(4)&0.286(4)&0.305(4)&0.321(4)&0.342(4)\\
 2.9&0.284(4)&0.291(4)&0.311(4)&0.327(4)&0.350(4)\\
 3.2&0.288(5)&0.296(4)&0.316(4)&0.333(4)&0.356(4)\\
 3.5&0.292(5)&0.300(5)&0.321(4)&0.338(4)&0.362(4)\\
 \hline
 \multicolumn{6}{c}{$K_{sea} = 0.1382$}\\
 \hline
 2.4&0.256(5)&0.262(5)&0.281(4)&0.298(4)&0.320(4)\\
 2.6&0.260(6)&0.266(5)&0.285(5)&0.302(4)&0.325(4)\\
 2.9&0.265(6)&0.271(5)&0.291(5)&0.309(5)&0.332(4)\\
 3.2&0.269(6)&0.276(6)&0.296(5)&0.315(5)&0.339(5)\\
 3.5&0.272(7)&0.280(6)&0.301(5)&0.320(5)&0.345(5)\\
 \hline
 \hline
 \multicolumn{6}{c}{$a^{3/2}(f\sqrt{M})^{(1)}$}\\
 \hline
 $aM_0$ &
 $K_{val}=0.1357$ & $K_{val}=0.1367$ & $K_{val}=0.1374$ &
 $K_{val}=0.1382$ & $K_{val}=0.1385$\\
 \hline
 \multicolumn{6}{c}{$K_{sea} = 0.1357$}\\
 \hline
 2.4&$-$0.0381(6)&$-$0.0387(6)&$-$0.0403(5)&$-$0.0416(5)&$-$0.0435(4)\\
 2.6&$-$0.0361(6)&$-$0.0367(5)&$-$0.0382(5)&$-$0.0395(5)&$-$0.0413(4)\\
 2.9&$-$0.0335(6)&$-$0.0341(5)&$-$0.0355(5)&$-$0.0367(4)&$-$0.0385(4)\\
 3.2&$-$0.0312(5)&$-$0.0318(5)&$-$0.0332(4)&$-$0.0343(4)&$-$0.0360(4)\\
 3.5&$-$0.0293(5)&$-$0.0298(4)&$-$0.0312(4)&$-$0.0322(4)&$-$0.0338(4)\\
 \hline
 \multicolumn{6}{c}{$K_{sea} = 0.1367$}\\
 \hline
 2.4&$-$0.0356(7)&$-$0.0363(6)&$-$0.0379(5)&$-$0.0393(5)&$-$0.0412(4)\\
 2.6&$-$0.0338(6)&$-$0.0344(6)&$-$0.0360(5)&$-$0.0373(5)&$-$0.0392(4)\\
 2.9&$-$0.0314(6)&$-$0.0319(6)&$-$0.0334(5)&$-$0.0347(5)&$-$0.0364(4)\\
 3.2&$-$0.0292(6)&$-$0.0298(5)&$-$0.0312(5)&$-$0.0324(4)&$-$0.0341(4)\\
 3.5&$-$0.0274(5)&$-$0.0279(5)&$-$0.0293(5)&$-$0.0304(4)&$-$0.0320(4)\\
 \hline
 \multicolumn{6}{c}{$K_{sea} = 0.1374$}\\
 \hline
 2.4&$-$0.0322(5)&$-$0.0329(5)&$-$0.0346(4)&$-$0.0361(4)&$-$0.0381(4)\\
 2.6&$-$0.0305(5)&$-$0.0311(5)&$-$0.0328(4)&$-$0.0342(4)&$-$0.0361(4)\\
 2.9&$-$0.0283(5)&$-$0.0288(4)&$-$0.0304(4)&$-$0.0318(4)&$-$0.0336(4)\\
 3.2&$-$0.0263(5)&$-$0.0268(4)&$-$0.0283(4)&$-$0.0296(4)&$-$0.0313(4)\\
 3.5&$-$0.0246(4)&$-$0.0251(4)&$-$0.0265(4)&$-$0.0277(3)&$-$0.0294(3)\\
 \hline
 \multicolumn{6}{c}{$K_{sea} = 0.1382$}\\
 \hline
 2.4&$-$0.0300(8)&$-$0.0306(7)&$-$0.0322(6)&$-$0.0337(5)&$-$0.0359(5)\\
 2.6&$-$0.0284(7)&$-$0.0289(6)&$-$0.0305(5)&$-$0.0320(5)&$-$0.0341(5)\\
 2.9&$-$0.0263(7)&$-$0.0268(6)&$-$0.0283(5)&$-$0.0297(5)&$-$0.0317(5)\\
 3.2&$-$0.0245(7)&$-$0.0250(6)&$-$0.0264(5)&$-$0.0278(5)&$-$0.0296(5)\\
 3.5&$-$0.0229(6)&$-$0.0234(6)&$-$0.0247(5)&$-$0.0261(5)&$-$0.0278(5)\\
 \end{tabular}
 \end{center}
\caption{Decay matrix elements at $\beta = 2.1$.}
\label{tab:frootm_beta2.1}
\end{table}

\begin{table}
  \begin{center}
    \begin{tabular}{cccccc}
      & $K$=0.1351& $K$=0.1365& $K$=0.1375& $K$=0.1385& $K$=0.1390\\
      \hline
      $aM_0$ & \multicolumn{5}{c}{$a^{3/2}(f\sqrt{M})^{(0)}$}\\
      \hline
      2.4&0.539(6)&0.548(6)&0.566(5)&0.583(5)&0.607(5)\\
      2.9&0.566(7)&0.575(6)&0.594(6)&0.613(6)&0.639(5)\\
      3.4&0.588(7)&0.598(7)&0.618(6)&0.638(6)&0.665(5)\\
      4.0&0.610(7)&0.621(7)&0.642(7)&0.663(6)&0.692(6)\\
      4.8&0.634(8)&0.646(7)&0.668(7)&0.690(7)&0.721(6)\\
      \hline
      \hline
      $aM_0$ & \multicolumn{5}{c}{$a^{3/2}(f\sqrt{M})^{(1)}$}\\
      \hline
      2.4&$-$0.0744(11)&$-$0.0748(10)&$-$0.0761(9)&$-$0.0772(9)&$-$0.0786(8)\\
      2.9&$-$0.0661(10)&$-$0.0665(10)&$-$0.0678(8)&$-$0.0689(7)&$-$0.0702(7)\\
      3.4&$-$0.0595(10)&$-$0.0599(9)&$-$0.0611(8)&$-$0.0622(7)&$-$0.0634(6)\\
      4.0&$-$0.0531(9)&$-$0.0535(8)&$-$0.0546(7)&$-$0.0557(6)&$-$0.0570(6)\\
      4.8&$-$0.0464(8)&$-$0.0468(8)&$-$0.0478(7)&$-$0.0488(6)&$-$0.0501(5)\\
    \end{tabular}
  \end{center}
  \caption{Decay matrix elements at $\beta = 2.187$.}
  \label{tab:frootm_beta2.187}
\end{table}
\begin{table}
  \begin{center}
    \begin{tabular}{cccccc}
      & $K$=0.1343& $K$=0.1357& $K$=0.1367& $K$=0.1377& $K$=0.1383\\
      \hline
      $aM_0$ & \multicolumn{5}{c}{$a^{3/2}(f\sqrt{M})^{(0)}$}\\
      \hline
      2.4&0.418(7)&0.430(8)&0.448(7)&0.466(6)&0.493(6)\\
      2.9&0.435(8)&0.448(8)&0.467(7)&0.486(7)&0.514(7)\\
      3.4&0.450(8)&0.463(9)&0.483(8)&0.502(7)&0.532(7)\\
      4.0&0.465(9)&0.478(10)&0.499(9)&0.519(8)&0.549(8)\\
      4.8&0.482(9)&0.495(11)&0.516(9)&0.537(9)&0.568(9)\\
      \hline
      \hline
      $aM_0$ & \multicolumn{5}{c}{$a^{3/2}(f\sqrt{M})^{(1)}$}\\
      \hline
      2.4&$-$0.0543(13)&$-$0.0549(11)&$-$0.0567(11)&$-$0.0580(10)&$-$0.0598(9)\\
      2.9&$-$0.0480(12)&$-$0.0486(10)&$-$0.0500(10)&$-$0.0513(9)&$-$0.0530(8)\\
      3.4&$-$0.0432(11)&$-$0.0437(9)&$-$0.0447(8)&$-$0.0459(9)&$-$0.0475(8)\\
      4.0&$-$0.0386(11)&$-$0.0390(9)&$-$0.0399(7)&$-$0.0408(8)&$-$0.0423(7)\\
      4.8&$-$0.0339(11)&$-$0.0342(9)&$-$0.0350(7)&$-$0.0356(8)&$-$0.0370(7)\\
    \end{tabular}
  \end{center}
  \caption{Decay matrix elements at $\beta = 2.281$.}
  \label{tab:frootm_beta2.281}
\end{table}
\begin{table}
  \begin{center}
    \begin{tabular}{cccccc}
      & $K$=0.1337& $K$=0.1349& $K$=0.1358& $K$=0.1368& $K$=0.1374\\
      \hline
      $aM_0$ & \multicolumn{5}{c}{$a^{3/2}(f\sqrt{M})^{(0)}$}\\
      \hline
      2.4&0.359(8)&0.370(6)&0.389(6)&0.406(5)&0.428(5)\\
      2.9&0.373(8)&0.385(7)&0.405(6)&0.423(5)&0.446(5)\\
      3.4&0.384(9)&0.396(7)&0.418(6)&0.437(6)&0.462(6)\\
      4.0&0.395(10)&0.408(8)&0.430(7)&0.451(6)&0.477(6)\\
      4.8&0.407(11)&0.421(9)&0.444(7)&0.466(7)&0.493(7)\\
      \hline
      \hline
      $aM_0$ & \multicolumn{5}{c}{$a^{3/2}(f\sqrt{M})^{(1)}$}\\
      \hline
      2.4&$-$0.0440(9)&$-$0.0449(8)&$-$0.0464(7)&$-$0.0477(6)&$-$0.0494(6)\\
      2.9&$-$0.0389(9)&$-$0.0396(7)&$-$0.0410(6)&$-$0.0423(6)&$-$0.0438(6)\\
      3.4&$-$0.0349(8)&$-$0.0356(7)&$-$0.0368(6)&$-$0.0380(6)&$-$0.0394(5)\\
      4.0&$-$0.0311(8)&$-$0.0317(6)&$-$0.0328(6)&$-$0.0339(5)&$-$0.0352(5)\\
      4.8&$-$0.0272(7)&$-$0.0277(6)&$-$0.0287(5)&$-$0.0296(5)&$-$0.0309(5)\\
    \end{tabular}
  \end{center}
  \caption{Decay matrix elements at $\beta = 2.334$.}
  \label{tab:frootm_beta2.334}
\end{table}
\begin{table}
  \begin{center}
    \begin{tabular}{cccccc}
      & $K=0.1329$ & $K=0.1337$ & $K=0.1344$ & $K=0.1351$ &
      $K=0.1353$\\ 
      \hline
      $aM_0$ & \multicolumn{5}{c}{$a^{3/2}(f\sqrt{M})^{(0)}$}\\
      \hline
      2.4&0.221(3)&0.224(3)&0.235(2)&0.247(2)&0.261(2)\\
      2.6&0.224(3)&0.227(3)&0.238(2)&0.251(2)&0.264(2)\\
      2.9&0.228(3)&0.231(3)&0.243(2)&0.255(2)&0.269(2)\\
      3.2&0.231(3)&0.234(3)&0.246(2)&0.259(2)&0.274(2)\\
      3.5&0.234(3)&0.237(3)&0.250(3)&0.263(2)&0.277(2)\\
      \hline
      \hline
      $aM_0$ & \multicolumn{5}{c}{$a^{3/2}(f\sqrt{M})^{(1)}$}\\
      \hline
      2.4&$-$0.0225(4)&$-$0.0227(3)&$-$0.0236(3)&$-$0.0245(3)&$-$0.0256(2)\\
      2.6&$-$0.0213(3)&$-$0.0215(3)&$-$0.0223(3)&$-$0.0232(3)&$-$0.0242(2)\\
      2.9&$-$0.0197(3)&$-$0.0199(3)&$-$0.0207(3)&$-$0.0215(2)&$-$0.0225(2)\\
      3.2&$-$0.0184(3)&$-$0.0185(3)&$-$0.0193(2)&$-$0.0200(2)&$-$0.0209(2)\\
      3.5&$-$0.0172(3)&$-$0.0173(3)&$-$0.0180(2)&$-$0.0188(2)&$-$0.0196(2)\\
    \end{tabular}
  \end{center}
  \caption{Decay matrix elements at $\beta = 2.575$.}
  \label{tab:frootm_beta2.575}
\end{table}


\newpage

\def\ie{{\sl i.e.}}
\def\etal{{\it et al.}}
\def\etc{{\it etc.}}
\def\ibid{{\it ibid}}


\newpage

\input epsf.tex 


%
%
%
%
%
%
%
%
%
%
%
%
%
%
%
%
%
%
%
%
\begin{figure}[tbp]  
  \begin{center}
    \epsfclipon
    \epsfxsize=0.6\hsize 
    \epsfbox{./fbsoverfb_nrqcd_quenched_vs_arho.eps}
    \epsfxsize=0.6\hsize 
    \epsfbox{./fbsoverfb_nrqcd_dynamical_vs_arho.eps}
  \end{center}
  \caption{
    $f_{B_s}/f_B$ as a function of the lattice spacing in quenched QCD (top)
    and full QCD (bottom). 
    Thick error bars 
    denote statistical errors, thin ones, systematic errors. 
    The solid horizontal line shows the final estimate for
    $f_{B_s}/f_B$ taken from the finest lattice, the dashed horizontal lines,
    the error on     this estimate.
    }  
  \label{fig:fbsoverfbscaling}
\end{figure}
%
%
%
%
%
%
%

\end{document}